\documentclass[twoside,twocolumn,9pt]{article}
\usepackage{extsizes}
\usepackage[super,sort&compress,comma]{natbib} 
\usepackage[left=1.5cm, right=1.5cm, top=1.785cm, bottom=2.0cm]{geometry}
\usepackage{balance}
\usepackage{times,mathptmx}
\usepackage{sectsty}
\usepackage{graphicx} 
\usepackage{lastpage}
\usepackage[format=plain,justification=justified,singlelinecheck=false,font={stretch=1.125,small,sf},labelfont=bf,labelsep=space]{caption}
\usepackage{float}
\usepackage{fancyhdr}
\usepackage{fnpos}
\usepackage[english]{babel}
\usepackage{array}
\usepackage{charter}
\usepackage[T1]{fontenc}
\usepackage[usenames,dvipsnames]{xcolor}
\usepackage{setspace}
\usepackage[compact]{titlesec}
\usepackage{tikz}

\usepackage[hyphens]{url}
\usepackage{amsmath}
\usepackage{amssymb}
\usepackage[caption=false]{subfig}
\usepackage{color}
\PassOptionsToPackage{caption=false}{subfig}
\usepackage{epsf}
\usepackage{sidecap}
\sidecaptionvpos{figure}{c}
\usepackage{amsmath,amssymb,bm,color,mathrsfs,wasysym}
\usepackage{epsfig}
\usepackage{amsfonts}
\usepackage{cancel}
\usepackage{ulem}
\usepackage{float}

\usepackage{tikz}
\usetikzlibrary{shapes}

\newcounter{saveeqn}

\newcommand{\be}{\begin{equation}}
\newcommand{\ee}{\end{equation}}
\newcommand{\bea}{\begin{eqnarray}}
\newcommand{\eea}{\end{eqnarray}}
\newcommand{\bi}{\begin{itemize}}
\newcommand{\ei}{\end{itemize}}

\newcommand{\etal}{\textit{et al.}}

\definecolor{LightBlue}{rgb}{0.0, 0.0, 1.0}

\newcommand{\tikzcircle}[2][black,fill=black]{\tikz[baseline=0ex]\draw[#1,radius=#2] (0,0.1) circle (0.1);}%
\newcommand{\sqdiamond}[1][fill=black]{\tikz [x=1.2ex,y=1.85ex,line width=.1ex,line join=round, yshift=-0.285ex] \draw  [#1]  (0,.5) -- (.6,1) -- (1.2,.5) -- (.6,0) -- (0,.5) -- cycle;}%
\newcommand{\MyDiamond}[1][fill=black]{\mathop{\raisebox{-0.275ex}{$\sqdiamond[#1]$}}}

\newcommand{\MySquare}[2][black,fill=black]{\tikz[baseline=0ex]\draw [fill=magenta,magenta] (0,0) rectangle (0.2,0.2);}%

\newcommand{\MyPentagon}{\raisebox{0pt}{\tikz{\node[draw,scale=0.7,regular polygon, regular polygon sides=5,fill=Maroon,rotate=0](){};}}}

\definecolor{darkgreen}{rgb}{0, 0.5, 0}
\definecolor{cream}{RGB}{222,217,201}

\begin{document}

\pagestyle{empty}


\makeFNbottom
\makeatletter
\renewcommand\LARGE{\@setfontsize\LARGE{15pt}{17}}
\renewcommand\Large{\@setfontsize\Large{12pt}{14}}
\renewcommand\large{\@setfontsize\large{10pt}{12}}
\renewcommand\footnotesize{\@setfontsize\footnotesize{7pt}{10}}
\makeatother

\renewcommand{\thefootnote}{\fnsymbol{footnote}}
\renewcommand\footnoterule{\vspace*{1pt}%
\color{cream}\hrule width 3.5in height 0.4pt \color{black}\vspace*{5pt}} 
\setcounter{secnumdepth}{5}

\makeatletter 
\renewcommand\@biblabel[1]{#1}            
\renewcommand\@makefntext[1]%
{\noindent\makebox[0pt][r]{\@thefnmark\,}#1}
\makeatother 
\renewcommand{\figurename}{\small{Fig.}~}
\sectionfont{\sffamily\Large}
\subsectionfont{\normalsize}
\subsubsectionfont{\bf}
\setstretch{1.125} 
\setlength{\skip\footins}{0.8cm}
\setlength{\footnotesep}{0.25cm}
\setlength{\jot}{10pt}
\titlespacing*{\section}{0pt}{4pt}{4pt}
\titlespacing*{\subsection}{0pt}{15pt}{1pt}


\makeatletter 
\newlength{\figrulesep} 
\setlength{\figrulesep}{0.5\textfloatsep} 

\newcommand{\topfigrule}{\vspace*{-1pt}%
\noindent{\color{cream}\rule[-\figrulesep]{\columnwidth}{1.5pt}} }

\newcommand{\botfigrule}{\vspace*{-2pt}%
\noindent{\color{cream}\rule[\figrulesep]{\columnwidth}{1.5pt}} }

\newcommand{\dblfigrule}{\vspace*{-1pt}%
\noindent{\color{cream}\rule[-\figrulesep]{\textwidth}{1.5pt}} }

\makeatother

\twocolumn[
  \begin{@twocolumnfalse}



\title{Adhesion dynamics of confined membranes}
\author{Tung B.T. To,\textit{$^{a, b}$} Thomas Le Goff,\textit{$^{a, c}$} and Olivier Pierre-Louis\textit{$^{a, \dag}$}}

\date{}

\maketitle

\begin{abstract}
We report on the modeling of the dynamics of confined lipid membranes.
We derive a thin film  model in the lubrication limit 
which describes an inextensible liquid membrane 
with bending rigidity confined between two adhesive walls. 
The resulting equations share similarities with the Swift-Hohenberg
model. However, inextensibility is enforced by a time-dependent nonlocal tension.
Depending on the excess membrane area available in the system, 
three different dynamical regimes, denoted as A, B and C, 
are found from the numerical solution of the model. 
In regime A, membranes with small excess area form flat adhesion domains and freeze.
Such freezing is interpreted by means
of an effective model for curvature-driven domain wall motion.
The nonlocal membrane tension tends to 
a negative value corresponding to the linear 
stability threshold of flat domain walls in the Swift-Hohenberg equation. 
In regime B, membranes with intermediate excess areas exhibit 
endless coarsening with coexistence of flat adhesion domains
and wrinkle domains.
The tension tends to the nonlinear 
stability threshold of flat domain walls in the Swift-Hohenberg equation. 
The fraction of the system covered by the wrinkle phase increases linearly with the
excess area in regime B.
In regime C, membranes with large  excess area 
are completely covered by a frozen labyrinthine pattern of wrinkles. 
As the excess area is increased, the tension increases and
the wavelength of the wrinkles decreases.
For large membrane area, there is a crossover to a regime where the extrema
of the wrinkles are in contact with the walls.
In all regimes after an initial transient, robust localised structures form,
leading to an exact conservation of the number of adhesion domains.
\end{abstract}

 \end{@twocolumnfalse} \vspace{0.6cm}
  ]

\renewcommand*\rmdefault{bch}\normalfont\upshape
\rmfamily
\section*{}
\vspace{-1cm}


\footnotetext{\textit{$^{a}$} Institut Lumi\`ere Mati\`ere, UMR5306 Universit\'e Lyon 1-CNRS, Universit\'e de Lyon, 69622 Villeurbanne, France}
\footnotetext{\textit{$^{b}$} Instituto de F\'{i}sica, Universidade Federal Fluminense, Avenida Litor\^{a}nea s/n, 24210-340 Niter\'{o}i, RJ, Brazil}
\footnotetext{\textit{$^{c}$} Aix Marseille Univ, CNRS, IBDM-UMR7288, 13009 Marseille, France}
\footnotetext{\textit{$^{\dag}$} Olivier.Pierre-Louis@univ-lyon1.fr}









\section{Introduction}
\label{sec:introduction}

Bilayer lipid membranes are abundant in biological systems \cite{Boal2002}. 
They are found in cell membranes, skin, eyes, articulations and pulmonary organs \cite{Hills2002,Hills20022,Das2009}.
Since their elasticity is dominated by bending rigidity,
lipid membranes exhibit specific morphologies and dynamics,
classifying their shape in a different class as compared to 
surface tension dominated phenomena, 
which govern the physics of capillarity and wetting.
Helfrich has first proposed a model energy functional within which
the behavior of lipid membranes could be explored~\cite{Helfrich1973}.
In the past two decades, much work has been devoted
to the analysis of the consequences of the Helfrich energy
on the morphology and dynamics of lipid membranes.
Various successes include the shape of vesicles
at equilibrium~\cite{Canham1970} or under hydrodynamic flow~\cite{Seifert1999,Kaoui2009}, or the
behavior of membrane stacks, and of supported
membranes on substrates~\cite{Radler1995,PierreLouis2008,Blachon2017}.

Here we wish to investigate  the consequences
of confinement on the dynamics
of membranes. We study the dynamics of a
fluid membrane with bending rigidity
and area conservation 
confined between two attractive walls. This geometry is firstly
motivated by the suggestion that membranes
can experience a double-well adhesion potential in cell adhesion processes \cite{Bruinsma2000},
or in biomimetic experiments~\cite{Sengupta2010}.
In these experiments, a short-range potential well 
located in the vicinity of the substrate results from the molecular
binding of ligand-receptor pairs. In addition, a free energy
barrier at intermediate ranges is provided by the entropic repulsion 
of a polymer brush
grafted to the substrate, which mimics the glycocalyx~\cite{Sackmann2014}. 
Finally, a long range attraction, resulting
either from Van der Waals forces \cite{Bruinsma2000}, 
or from gravity \cite{Sengupta2010}
enforces a second potential well for the membrane.
In parallel with these works, other studies in the literature 
allow one to formulate a different picture for membrane adhesion,
which is also based on the concept of a double-well potential.
Indeed, blebbing \cite{Brugues2010,Charras2008,Paluch2005} ---the local detachment of cell membranes 
from the cytoskeleton, suggests that the link between
membranes and the cytoskeleton can be described to some
extent by simple adhesion concepts. Cell adhesion
could therefore be mimicked by a competition
between this adhesion of the membrane to the cytoskeleton, 
and adhesion to a substrate.
Such a scenario is reminiscent of that proposed in Ref. \cite{Speck2012},
where membranes experience competing adhesion between  a
substrate and the cytoskeleton.
Moreover, a third and somewhat different 
situation involving a double-well potential arises when two types ligand-receptor
pairs with different lengths compete for adhesion, 
leading to two possible equilibrium separation distances,
as proposed in Ref. \cite{Asfaw2006}.

Our study is inspired by these diverse pictures of membrane adhesion
in two-state, or double-well potentials.
Our aim here is to capture some generic features 
of these systems by exploring the simple case of a membrane
confined between two flat adhesive
walls. Adhesion in biological cells 
involving e.g. signaling, the remodeling of the cytoskeleton,
or the diffusion and clustering of ligands and receptors,
is certainly more complex than our minimal modeling approach.
However, we hope that our results will provide hints
to understand systems involving more physical ingredients.

On a more fundamental and theoretical level, 
our model for membrane dynamics defines
a novel universality class for phase separation
in two dimensions with unique features. 
Indeed, standard models for phase separation
were developed to study spinodal decomposition in alloys \cite{Langer1971}, 
binary fluids \cite{Siggia1979}, reaction-diffusion, 
magnetism and wetting phenomena. 
They are generically described by the time-dependent Ginzburg-Landau 
(TDGL) equation --also called the Cahn-Allen equation \cite{Hohenberg1977},
or its conserved  version the Cahn-Hilliard equation \cite{CahnHilliard1958},
and give rise to power-law coarsening (i.e. perpetual 
increase of the domain size) via curvature-driven
motion of domain walls. Here, we show that
the dynamical equations governing confined membranes 
share similarities with the Swift-Hohenberg equation,
however with a time-dependent tension that enforces membrane area conservation.
Within this model, membrane adhesion domains exhibit
a transition to coarsening controlled
by the total excess area of the membrane.

The results reported here build on
our previous study of membrane adhesion dynamics 
based on a one-dimensional (1D) model with bending rigidity, but without area conservation \cite{LeGoff2014,LeGoff2015StatMech,LeGoff2015PRE}.
In 1D, we found that bending rigidity induces  oscillatory
interactions between domain walls. As a consequence of these oscillations, the dynamics freezes into a disordered 
or ordered profile depending on the permeability of the walls. 
In contrast to this behavior, we show here that in 
two-dimensions (2D) and without area conservation 
these oscillatory interactions between domain walls
do not affect the coarsening behavior. Indeed,
we recover standard coarsening with
the same exponents as that of usual phase separation models
(TDGL or Cahn-Hilliard).
The study of such a model without area conservation can be motivated by
the investigation of the coarsening dynamics of
2D systems at the Lifshitz-point, defined as
the point where the prefactor of the gradient-squared
term in the Landau free energy density vanishes~\cite{Hornreich1975},
and where higher-order bending-like squared-Laplacian terms come into play.

However, the present study focuses on the case
of 2D membranes where local area conservation should be imposed.
As discussed above, this leads to the presence of a time-dependent tension in the dynamical
equations. 
The numerical solution of these equations reveals three regimes,
hereafter denoted as A, B and C,
depending on the excess area of the membrane.

For small excess area (regime A), the membrane freezes into
a state with flat adhesion patches of finite size.
This freezing can be understood as follows.
Domain wall motion is driven by an
effective positive wall tension, 
and acts so as to reduces the total domain wall length.
Due to area conservation, domain wall length decrease
implies an increase of the membrane excess area per unit length
in domain walls. This increase
induces a cancellation
of the domain wall tension, leading to
the arrest of domain wall motion.
The dynamics in regime A can be analyzed within
a simple model for the coupled dynamics
of domain wall motion
and of the nonlocal membrane tension.
Since it corresponds to the cancellation
of wall tension, the asymptotic value of the nonlocal tension
is equal to the threshold tension for linear instability
of domain walls in the SH equation.

For intermediate excess area (regime B),
the membrane exhibits coarsening with a coexistence
of flat and wrinkled domains. The wrinkle phase
forms spontaneously when domain walls collide.
The fraction of the 
system occupied by the wrinkle phase 
reaches a constant value at long times, which 
increases linearly with the
excess area. Concurrently, the size of the 
flat and wrinkled domains increases indefinitely with time.
Since the system reaches coexistence of wrinkles and
flat domains, the asymptotic value of the nonlocal tension
corresponds to the threshold for nonlinear
stability of domain walls in the SH equation.

For larger excess area (regime C), 
the wrinkle phase invades the whole system and the membrane freezes into a labyrinthine wrinkle pattern.
The amplitude and wavelength of the wrinkles
are analyzed within a simple sinusoidal ansatz,
that reveals the presence of two different regimes within regime C.
For large excess area the amplitude of the wrinkles
is fixed by the contact with the two walls,
while wrinkles with smaller excess area 
exhibit a free amplitude smaller than the distance
between the two walls.

We focus on the case with permeable walls, but also briefly
report on the behavior of a simplified model 
for impermeable walls which exhibits similar dynamics.
We also find unexpectedly that the number of adhesion
domains is strictly conserved in all cases during 
the late stages of the dynamics due 
to the formation of robust localised structures
which forbid the complete disappearance of the adhesion domains.





\section{Model Equations}
\label{s:model}


\begin{figure}[ht]
    \begin{center}
            \includegraphics[width=0.48\textwidth]{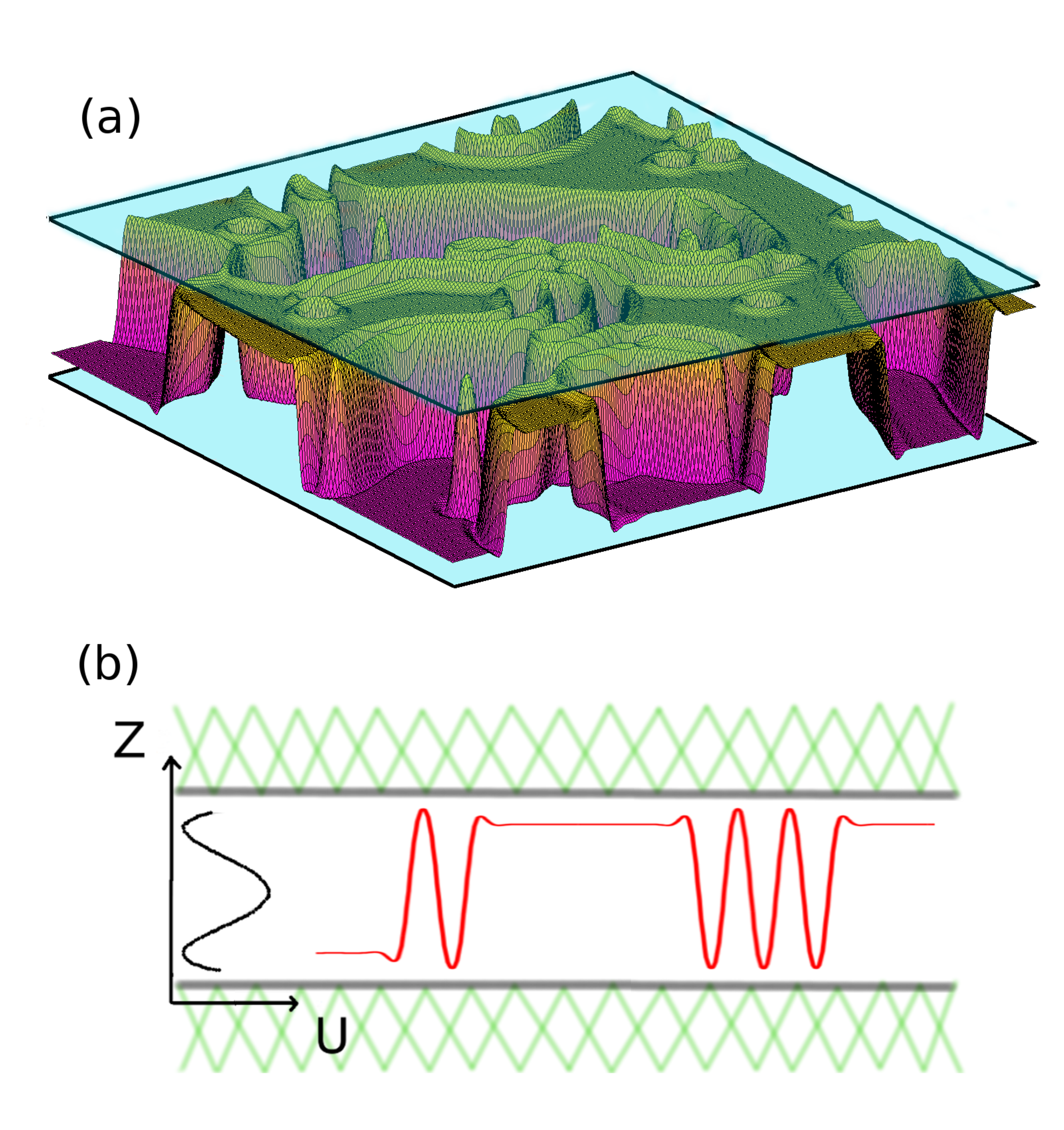}
    \caption{
(Color online) Membrane between two adhesive substrates. 
(a) 3D view; (b) 2D schematics: left black curve: double-well potential, right red curve: a slice of the membrane profile (profiles are obtained from simulation with  
$\Delta A^*=3.61 \cdot 10^{-2}$ at $T=8 \cdot 10^5$, 
corresponding to the last figure in regime B2 of Fig. \ref{fig:figure2}). 
\label{fig:figure1}
		}
	\end{center}
\end{figure}

A schematic representation of a membrane of height $h(x,y,t)$ along $z$ confined between two parallel flat walls located at $z=\pm h_0$ is shown in Fig. \ref{fig:figure1}(a,b).
The interaction between the membrane and the walls 
is modeled via a double-well potential ${\cal U}(h)$. 
Adding this interaction energy with the Helfrich bending energy,
we obtain the total energy of the membrane as
\begin{eqnarray}
{\cal E} = \int d {\cal A} \Big( \frac{\kappa}{2} {\cal C}^2 + {\cal U}(h) \Big),
\label{eq:membraneenergy}
\end{eqnarray}
where $d{\cal A}$ denotes the infinitesimal area element of the membrane,
${\cal C}$ denotes the membrane local curvature 
and $\kappa$ denotes the bending rigidity. 

In addition, the local conservation of the membrane area can be expressed
as a conservation of the membrane area density 
$\rho(x,y,t)=[1+(\nabla_{xy} h)^2]^{1/2}$:
\begin{eqnarray}
\partial_t \rho + \nabla_{xy} . (\rho \mathbf{v}_{xy})=0,
\label{e:rho_2D}
\end{eqnarray}
where $\mathbf{v}_{xy}$ is the membrane velocity parallel 
to the walls, $\nabla_{xy}$ is the gradient in the (x,y) plane.
In order to enforce this constraint, 
we make use of a local space-dependent 
and time-dependent Lagrange multiplier $\sigma(x,y,t)$
(which can be interpreted as a local
membrane tension \cite{Campelo2006,Kaoui2008}).
The membrane energy is thus generalised as
\begin{eqnarray}
{\cal X} = 
\int d {\cal A} \Big( \frac{\kappa}{2} {\cal C}^2 + {\cal U}(h) +\sigma\Big).
\label{eq:membrane_Xi}
\end{eqnarray}
The local tension $\sigma(x,y,t)$ 
leads to an additional contribution to membrane forces,
which is constrained to obey Eq. (\ref{e:rho_2D}) at all times.
Such a constraint allows one to determine $\sigma$.

Since all the phenomena described here occur at 
small scales, we consider the Stokes regime where inertia is negligible.
Hence, membrane forces $\mathbf f$ resulting from energy variations and
inextensibility have to balance viscous forces
exerted by the surrounding liquid on the membrane: 
\begin{eqnarray}
\mathbf f=(\mathbf s_+|_{z=h(x,y,t)}-\mathbf  s_-|_{z=h(x,y,t)}) \cdot \mathbf n
\label{e:membrane_force_equilibrium}
\end{eqnarray}
where $\mathbf s$ is the stress tensor in the liquid, $\pm$ denotes
the liquid above or below the membrane, and $\mathbf n$
is the normal to the membrane. 
The liquid with velocity $\mathbf v_\pm$ obeys
the incompressible Stokes equations, 
with $\nabla \cdot \mathbf v=0$
and  $\nabla \cdot \mathbf s=0$, 
where $\mathbf s_{ij}=\mu (\partial_iv_j+\partial_jv_i)-p\delta_{ji}$ where $\mu$ is the fluid viscosity and $p$ its pressure.
In addition, we assume no-slip~\cite{Mueller2009,Botan2015}  and 
impermeability~\cite{Huster1997,Fettiplace1980,Mathai2008}
at the membrane, leading to
\begin{align}
\mathbf{v}_+|_{z=h(x,y,t)}=\mathbf{v}_-|_{z=h(x,y,t)}.
\label{e:kinematic_membrane}
\end{align}
In order to account for the permeability
of the walls, we impose 
\begin{eqnarray}
v_{z}|_{z=\pm h_0}=\pm \nu(p_\pm-p_{ext}),
\label{e:BC_nu}
\end{eqnarray}
 where  $p_{ext}$ is a reference constant pressure,
and $\nu$ is the wall permeability.
Finally, 
we assume a no-slip condition at the walls $\mathbf v_{xy\pm}|_{z=\pm h_0}=0$.

\section{Lubrication limit: permeable walls}
\label{s:lubricationpermeable}

We apply the lubrication limit 
where $|\nabla_{xy} h|$ is small. In this limit,
the hydrodynamic flow above and below the membrane
are simple Poiseuille flows parallel to the walls.
Using the boundary conditions presented in the previous section,
the hydrodynamic flows can be obtained explicitly.
We then obtain dynamical equations for the membrane profile
using Eq. (\ref{e:kinematic_membrane}).

The derivation of the evolution equation for $h$ in this limit
actually follows the same steps as the simpler case
of a one-dimensional membrane without area conservation
reported in Ref. \cite{LeGoff2014}.
The main difference comes from
the constraint of membrane incompressibility.
The details of the calculations 
and the general result are reported in Appendix \ref{a:lubrication}. 
We will mainly focus on the limit
of large normalised permeability, $\bar\nu\gg 1$, with
\begin{align}
\bar{\nu}=\frac{12\mu\nu \kappa^{1/2}}{h_0^2 {\cal U}_0^{1/2}}.
\end{align}
where ${\cal U}_0$ is the amplitude 
of the interaction potential.

In physical systems, the value of $\bar\nu$
depends crucially on the permeability $\nu$ of the substrate. 
One possibility to evaluate $\bar\nu$ 
is to use Darcy's law for porous media,
using~\cite{Richardson1972} $\nu\sim a_0^2/(\mu h_w)$, 
where $a_0$ is the scale of the pores and $h_w$ the 
thickness of the wall. 
This suggests $\bar\nu\sim 12a_0^2\kappa^{1/2}/(h_0^2h_w{\cal U}_0^{1/2})$.
In the case of scaffolded actin cytoskeleton, the typical pore size is $a_0\sim 10^{-7}$m and the thickness $h_w \sim 10^{-6}$m. 
Moreover, assuming that the energy-scale of the potential is dictated by 
the cytoskeleton-lipid membrane adhesion energy, we find
\cite{Raucher2000} ${\cal U}_0\sim 10^{-5}J.m^{-2}$.
Using~\cite{Henriksen2006, Pakkanen2011}
$\kappa\sim10^{-19}J$ and 
$h_0\sim10^{-8}m$, 
we find $\bar{\nu}\sim10^2$. 
For collagen, a very common extracellular matrix, 
the typical pore size is $10^{-6}$m and thickness is $10^{-4}$m \cite{Miron2010} which gives same order of magnitude for $\bar{\nu}$.

However, if the substrate is covered by another lipid membrane,
the permeability of the substrate becomes very small 
~\cite{Huster1997,Fettiplace1980,Mathai2008}.
An increase of about 1 atm of osmotic pressure induces a speed of about $10^{-4}$m.s$^{-1}$ for the water across a lipid membrane~\cite{Huster1997,Fettiplace1980,Mathai2008}. This means that from Eq. \eqref{e:BC_nu} $\nu\sim10^{-9}$m$^2$.s.kg$^{-1}$. Then using water viscosity,
and the same values as above $\kappa\sim10^{-19}J$, $h_0\sim10^{-8}$m and ${\cal U}_0\sim10^{-5}$J.m$^{-2}$, we find $\bar{\nu}\sim10^{-2}$.
We will briefly discuss the case of small permeabilities
in the end of the paper.

In the limit of strongly permeable walls $\bar{\nu} \gg 1$,
the dynamical equation reads
\begin{eqnarray}
\partial _t h = \frac{\nu}{2}(- \kappa \Delta ^2 h + \sigma_0 \Delta h - {\cal U} ^\prime (h)),
\label{eq:hpermeable}
\\
\sigma_0 = \frac{\int \int dx dy \Big(\kappa \Delta ^2 h + {\cal U} ^\prime (h)\Big)\Delta h}{\int \int dx dy (\Delta h)^2}.
\label{eq:sigmapermeable}
\end{eqnarray}
Here, $\Delta$ denotes the Laplacian operator in  planar coordinates $(x,y)$. 
One important and non-trivial 
result emerging from the lubrication limit is that,
to leading order, the local Lagrange multiplier $\sigma(x,y,t)$ is constant in space, 
with a value $\sigma_0(t)$,
thereby leading to a nonlocal area conservation constraint.

The evolution equations Eqs. (\ref{eq:hpermeable},\ref{eq:sigmapermeable})
decrease the energy, here written in the small slope limit as
\begin{eqnarray}
{\cal E} = \int d {\cal A} \Big( \frac{\kappa}{2} (\nabla^2h)^2 + {\cal U}(h) \Big).
\label{eq:membrane_E}
\end{eqnarray}
Indeed, as shown in Appendix \ref{b:energyrelaxation} 
using the Schwarz inequality, 
we always have $\partial_t{\cal E}\leq 0$.

\section{Normalization and numerical methods}

In order to discuss the dynamical behavior of Eqs. (\ref{eq:hpermeable},\ref{eq:sigmapermeable})
and to solve the equations numerically,
we normalise space and time, leading to
\begin{eqnarray}
\partial _T H = -  \Delta ^2 H + \Sigma \Delta H -  U'(H),
\label{eq:hpermeable_norm}
\\
\Sigma = \frac{\int \int dX dY \Big( \Delta ^2 H+  U' (h)\Big)\Delta H}
{\int \int dX dY (\Delta H)^2}.
\label{eq:sigmapermeable_norm}
\end{eqnarray}
where $U(h)={\cal U}(H)/{\cal U}_0$,  $T=\nu{\cal U}_0t/(2h_0^2)$, 
$H=h/h_0$, $X=x/{\cal L}_0$, $Y=y/{\cal L}_0$,
with ${\cal L}_0=\kappa^{1/4} h_0^{1/2}/{\cal U}_0^{1/4}$,
and $\sigma_0=\Sigma ({\cal U}_0\kappa)^{1/2}/h_0$.
Using the numerical values of the previous section,
we obtain the order of magnitude of tensions $({\cal U}_0\kappa)^{1/2}/h_0\sim 10^{-4}$J.m$^{-2}$, 
and the typical lengthscale parallel to the membrane ${\cal L}_0\sim 30$nm. 
The small slope limit amounts to considering $h_0/{\cal L}_0\approx 0.3$ is small as compared to 1.
Although these slopes are not very small, the lubrication limit
is known to provide a qualitatively good and robust description of the physical behavior
for moderate slopes~\cite{Snoeijer2006}.
In addition, smaller adhesion energies, e.g., using ${\cal U}_0\sim 2\times 10^{-6}$Jm$^{-2}$ as
suggested by Ref.\cite{Charras2008},
or ${\cal U}_0\sim 1.4\times 10^{-6}$Jm$^{-2}$
in Ref.\cite{Sengupta2010},
lead to even smaller slopes.
In general, the adhesion energy, on which the validity of this limit
depends crucially, is system-dependent and varies strongly with the type and the density of binders.
Considering the different case of physical adhesion with a 
porous solid substrate,
a crude approximation consists in multiplying the 
physical adhesion potential of Ref.\cite{Swain2001}
---based on Van der Waals and hydration interactions,
with the solid fraction $\vartheta$. Assuming for example
$\vartheta\approx 0.3$, we then find a value
${\cal U}_0\approx 1.5\times 10^{-6}$Jm$^{-2}$
similar to those reported above.
Summarizing this discussion, 
we expect in most cases ${\cal L}_0\sim 30$ to $50$nm
for nano-confinement with $h_0\sim 10$nm.

We choose a specific form of the double-well adhesion potential
\begin{eqnarray}
U(H) = \frac{1}{4} \left( H_m^2 - H^2 \right)^2
+U_d 
\end{eqnarray}
where 
\begin{align}
U_d= U_1[{\rm e}^{-(1-H)/d}+{\rm e}^{-(1+H)/d}]
\label{e:shortrange_Ud}
\end{align}
is a short-range repulsion, the aim of which is
to avoid collision of the membrane with the walls.
In the simulations, we have chosen $H_m=0.7$, $d=0.01$.
Assuming that $h_0\sim 10$nm, the position of the minimum of the potential corresponds to 
a distance to the substrate $h_0(1-H_m)\sim 3$nm,
which is similar to those reported in the literature
for physical interactions\cite{Swain2001},
or binders \cite{Sengupta2010}.
During the simulations, we usually assumed that $U_1=1$. However,
we used $U_1=0$ to accelerate long simulations 
in the regimes where the membrane
did not approach the walls.

The dynamics is integrated by means of  
a first-order exponential-time-differencing method (ETD1) in Fourier space \cite{CoxMatthews2002} with space and time discretization bins $dX=0.4$, and $dT=0.15$.
We have modified the integration scheme
in order to conserve exactly the excess area 
\begin{eqnarray}
\Delta  A = \frac{1}{2} \int \int dX dY (\nabla H)^2.
\label{e:excess_area_normalised}
\end{eqnarray}
Our scheme amounts to choosing
a specific discretization of the expression of the tension $\Sigma$ 
in Eq. (\ref{eq:sigmapermeable_norm}) in order to enforce exact area conservation.
The details of this scheme are presented in Appendix \ref{a:area_conservation}.

The simulation box sizes where usually
$L_X\times L_Y=400\times 400$, or $800\times 800$ corresponding
to physical sizes from $10$ to $40\mu$m.




\section{Area-preserving vs tensionless membranes}
\label{sec:results}

\begin{figure*}[htb!]
    \begin{center}
            \includegraphics[width=0.6\textwidth]{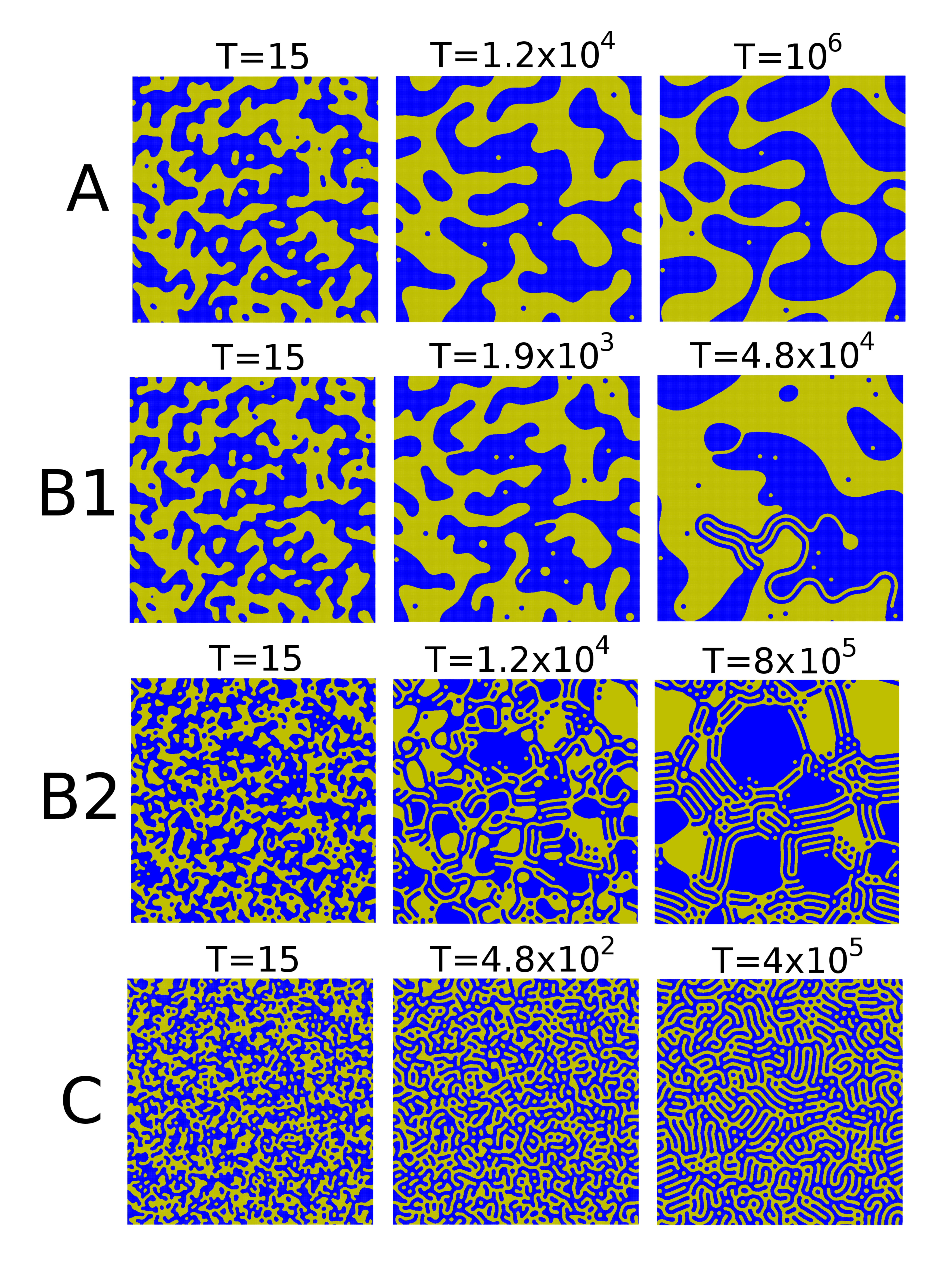}
    \caption{
(Color online) Membrane adhesion dynamics for various excess areas. Yellow: adhesion patches
 on the upper wall. Dark blue: adhesion patches on the lower wall. Regime A with freezing
  of flat domains for small excess area : $\Delta A^*=0.88 \times 10^{-2}$.  
Regime B with coexistence of the flat-domain phase and wrinkle phase
 with coarsening for intermediate excess area: B1 with $\Delta A^*=1.08 \times 10^{-2}$,
 and B2 with $\Delta A^*=3.61 \times 10^{-2}$. 
 Regime C with frozen wrinkles for larger excess area $\Delta A^*=5.68 \times 10^{-2}$. 
 System size $L \times L=400\times400$.	}
        \label{fig:figure2}
	\end{center}
\end{figure*}

We start with noisy initial conditions (details on these
conditions are described in Appendix \ref{a:initial_conditions}).
Three different regimes are obtained depending on the 
 excess area density
\begin{align}
\Delta A^*=\frac{\Delta A}{A_{syst}}.
\end{align}
where  $A_{syst}$ is the system size in the $(X,Y)$ plane
(if the simulations are performed in a rectangular 
box of size $L_X\times L_Y$, we have $A_{syst}=L_XL_Y$).
For increasing excess area density, we first find a
regime with frozen flat domains, then a regime with
coarsening and with coexistence between flat domains and wrinkles,
and finally a regime with a disordered pattern of frozen wrinkles.
These three regimes will be
denoted as regimes A, B and C in the following.
Some snapshots of these evolutions are shown in Fig.~\ref{fig:figure2}.

\begin{figure*}[ht]
    \begin{center}
    \subfloat[]{\includegraphics[width=0.6\textwidth]{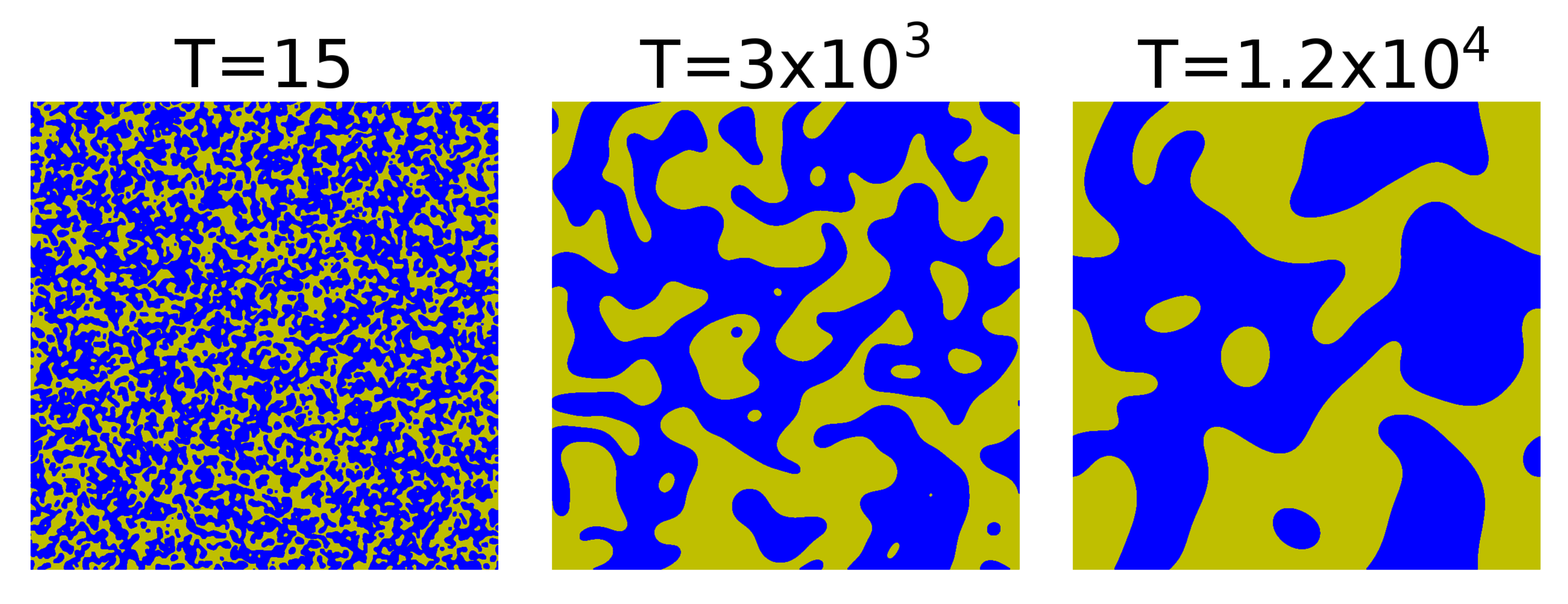}} \\
    \subfloat[]{\includegraphics[width=0.6\textwidth]{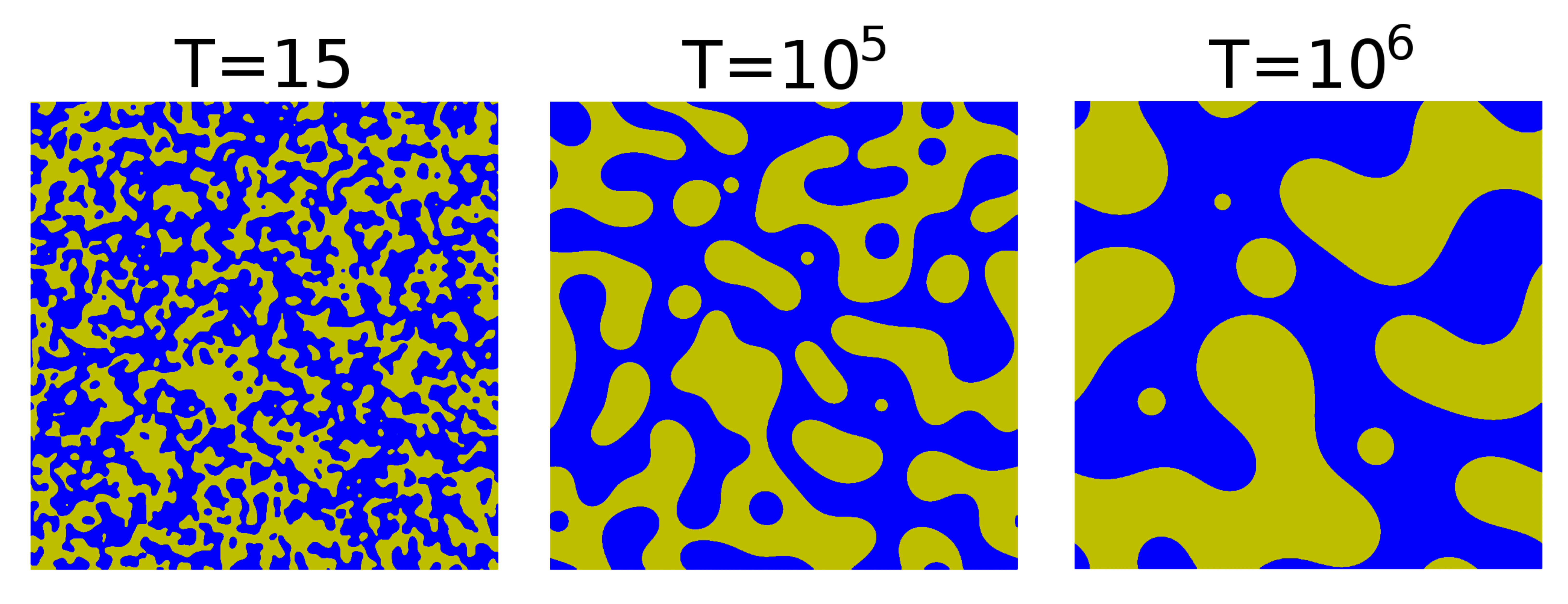}}
    \caption{
(Color online) (a) TDGL4 (Eq. (\ref{eq:hpermeable}) with $\Sigma=0$), $L=1600$, $L_{DW} \sim T^{1/2}$. (b) CH4 (constant mobility) (Eq. (\ref{eq:dhdt_impermeable}) with $M(H)=1$ and $\Sigma=0$), $L=800$, $L_{DW} \sim T^{1/3}$.}
        \label{fig:tdgl4ch4cm}
	\end{center}
\end{figure*}
Since frozen states were also observed in 
a one-dimensional model without area conservation in Ref. \cite{LeGoff2014},
we wish to investigate the behavior of the model
without the constraint of area conservation as a 
preamble to the analysis of the full model. This is done by 
simulating Eq. (\ref{eq:hpermeable_norm}) with $\Sigma=0$.
The resulting equation was called the fourth order Time-dependent
Ginzburg-Landau equation (TDGL4) in Ref. \cite{LeGoff2014}.
Such an equation corresponds to a system
that would exhibit bending rigidity, but for which
the extension of the area of the interface would occur at no cost
(physically corresponding to dynamics at a Lifshitz point).

Frozen disordered states obtained in 
simulations of TDGL4 in 1D~\cite{LeGoff2014} can
be seen as a consequence of trapping of domain-walls (called kinks in 1D)
into their mutual oscillatory interactions.
These oscillations, which can be traced back to the oscillatory tails
of the domain wall profiles, are still present in 2D. 
The first oscillation can indeed be observed
in the vicinity of all domain walls
in Fig. \ref{fig:figure1}.
However, the TDGL4 dynamics in 2D,
shown in Fig. \ref{fig:tdgl4ch4cm}(a), actually leads to a simple coarsening behavior
with flat adhesion patches, the size of which grows like $t^{1/2}$.
This is identical to the 2D behavior of the TDGL equation.
Such a coarsening behavior is usually interpreted as a consequence of motion 
of domain walls driven by their curvature. Thus, the coarsening
behavior of TDGL4 suggests that, in absence of area-conservation constraint,
motion by curvature of domain walls dominates over oscillatory interactions.

As mentioned above, membranes with area conservation
exhibit strikingly different dynamics. Hereafter, we discuss 
the three dynamical regimes arising in membranes with area 
conservation.

\section{Small excess area}
\label{s:regimeA}

\begin{figure}[ht]
    \begin{center}
            \includegraphics[width=0.5\textwidth]{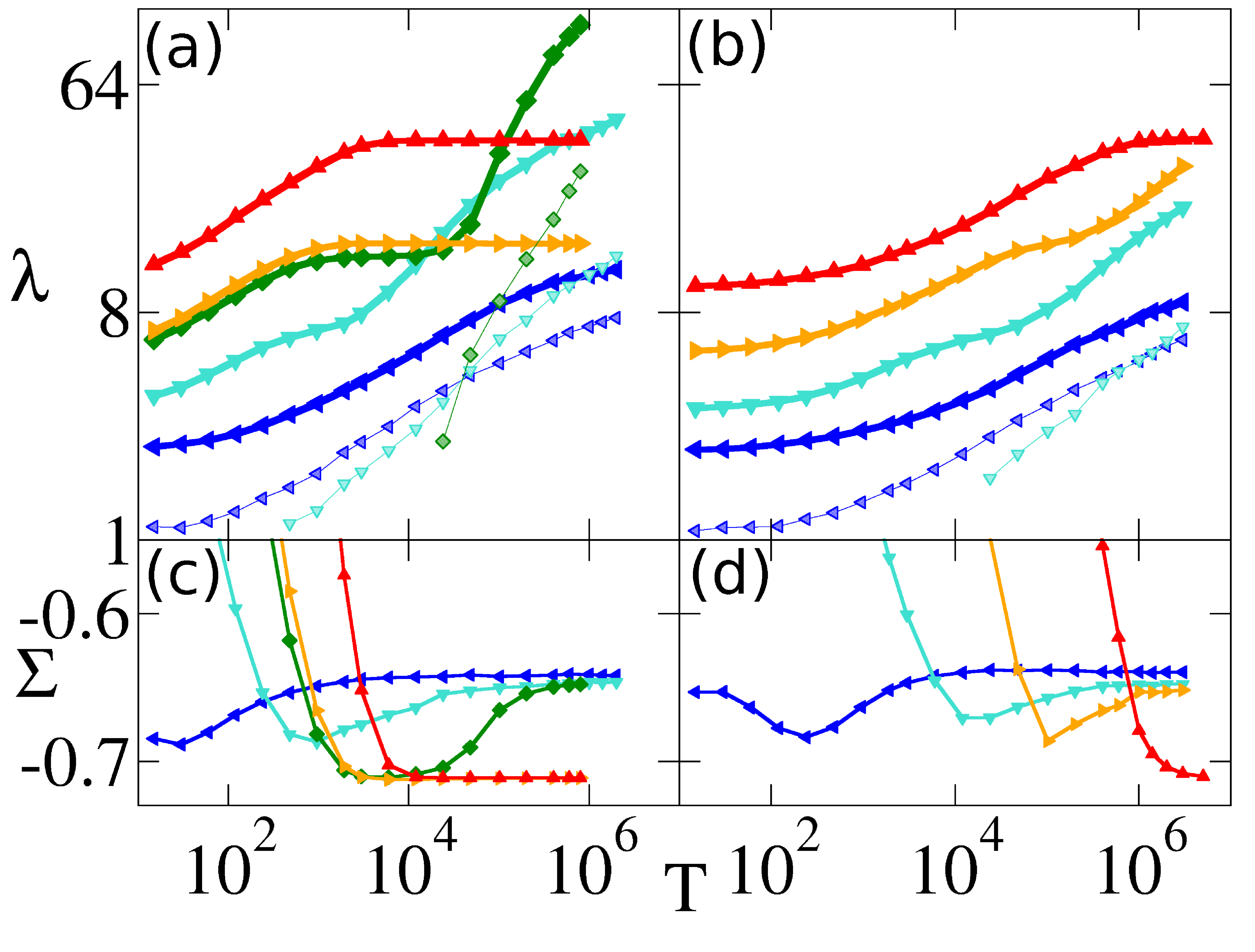}
    \caption{
(Color online) Time-evolution of the lengthscale $\lambda$ and the tension $\Sigma$.
Size of the simulation box $L=800$. 
(a) and (c): results in the limit  of large wall permeability. 
(b) and (d): results for impermeable walls. 
In (a) and (b), thick lines correspond to $\lambda _{flat}$,
and thin lines to $\lambda _{wr}$. Excess area:
({\color{red} $\blacktriangle$}) $\Delta A^*=0.37 \times 10^{-2}$;
({\color{orange} $\blacktriangleright$}) $\Delta A^*=0.88 \times 10^{-2}$;
($\MyDiamond[draw=OliveGreen,fill=OliveGreen]$) $\Delta A^*=0.96 \times 10^{-2}$;
({\color{BlueGreen} $\blacktriangledown$}) $\Delta A^*=1.8 \times 10^{-2}$;
({\color{LightBlue} $\blacktriangleleft$}) $\Delta A^*=3.61 \times 10^{-2}$.
}
\label{fig:figure3}
	\end{center}
\end{figure}

Regime A corresponds to simulations of 
Eqs. (\ref{eq:hpermeable_norm},\ref{eq:sigmapermeable_norm})
at small excess area. 
In this regime, flat adhesion domains expand initially (Fig. \ref{fig:figure2}A, $T=15$),
and the tension $\Sigma$ decreases to negative values.
Later, these domains freeze (\ref{fig:figure2}A, $T=10^6$), and $\Sigma$ reaches
a constant negative value independent of initial conditions and of the imposed
excess area. The evolution of the domain size $\lambda_{flat}$ 
and of the tension are shown in Fig. \ref{fig:figure3}(a,c).


When $\Sigma$ is constant in time,
our dynamical equation (\ref{eq:hpermeable_norm}) is identical to the much-studied Swift-Hohenberg (SH) equation \cite{Elder1992,Cross1995,Ouchi1996,LeBerre2002,Hagberg2006}
(neglecting the contribution $U_d$ to the potential).
Hence, the steady-states of our model are steady-states 
of the SH equation. However, the stability of these steady-states can be different.
We observe that the value towards which  $\Sigma$ converges in our simulations
is close to the  limit of linear stability of flat domain walls with 
respect to transverse perturbation in the SH model at $\Sigma_c \approx -1.0226 H_m$ 
as reported by Hagberg \etal \cite{Hagberg2006}. 
Such a limit of stability is associated with the cancellation
of the energy of domain walls in the SH equation.

In order to explore the consequence of this cancellation,
we need to relate the energy in the two models.
In normalised form, they read:
\begin{align}
E&=\int dA
\left\{
\frac{1}{2}(\Delta H)^2 + U(H) 
\right\},
\label{e:E_energy_rescaled}
\\
\Xi & = \int dA 
\left\{
\frac{1}{2} (\Delta H)^2 + U(H) 
+\frac{\Sigma}{2} (\nabla H)^2
\right\}.
\label{e:Xi_energy_rescaled}
\end{align} 

The energies $e_{DW}$ and $\xi_{DW}$ per unit length of a 
straight and isolated domain wall in our model and in the SH model therefore read
\begin{align}
e_{DW}&=\int d\zeta 
\left\{
\frac{1}{2}(\partial_{\zeta \zeta} H)^2 + U(H) 
\right\},
\label{e:E_wall}
\\
\xi_{DW} &= \int d\zeta 
\left\{
\frac{1}{2} (\partial_{\zeta \zeta} H)^2 + U(H) 
+\frac{\Sigma}{2} (\partial_\zeta H)^2
\right\},
\label{e:xi_wall}
\end{align} 
where $\zeta$ is the coordinate orthogonal to the domain wall.
Assuming that the whole excess area is stored in domain walls,  
we have
\begin{align}
\Delta A \approx L_{DW}\alpha_{DW}.
\label{eq:areazeta}
\end{align} 
where $L_{DW}$ is the length of domain walls in the system,
and 
\begin{align}
\alpha_{DW}=\frac{1}{2}\int d\zeta (\partial _\zeta H)^2
\end{align}
is the excess area per unit domain wall length.
Combining Eqs. (\ref{e:E_wall},\ref{e:xi_wall},\ref{eq:areazeta}), we find
\begin{align}
\xi_{DW} = e_{DW} +  \frac{\Sigma \Delta A}{ L_{DW}}. 
\end{align} 
Since $\xi_{DW}=0$ when $\Sigma=\Sigma_c$, we obtain an expression
for the size of the domains in the frozen state of our model
\begin{align}
\bar{\lambda}_{flat}
=\frac{A_{syst}}{L_{DW}}
=\frac{e_{DW}^c}{-\Sigma_c \Delta A^*},
\label{e:bar_lambda_flat}
\end{align} 
where $e_{DW}^c$  is the value of $e_{DW}$
for $\Sigma=\Sigma_c$. 

\begin{figure}
    \begin{center}
            \includegraphics[width=0.5\textwidth]{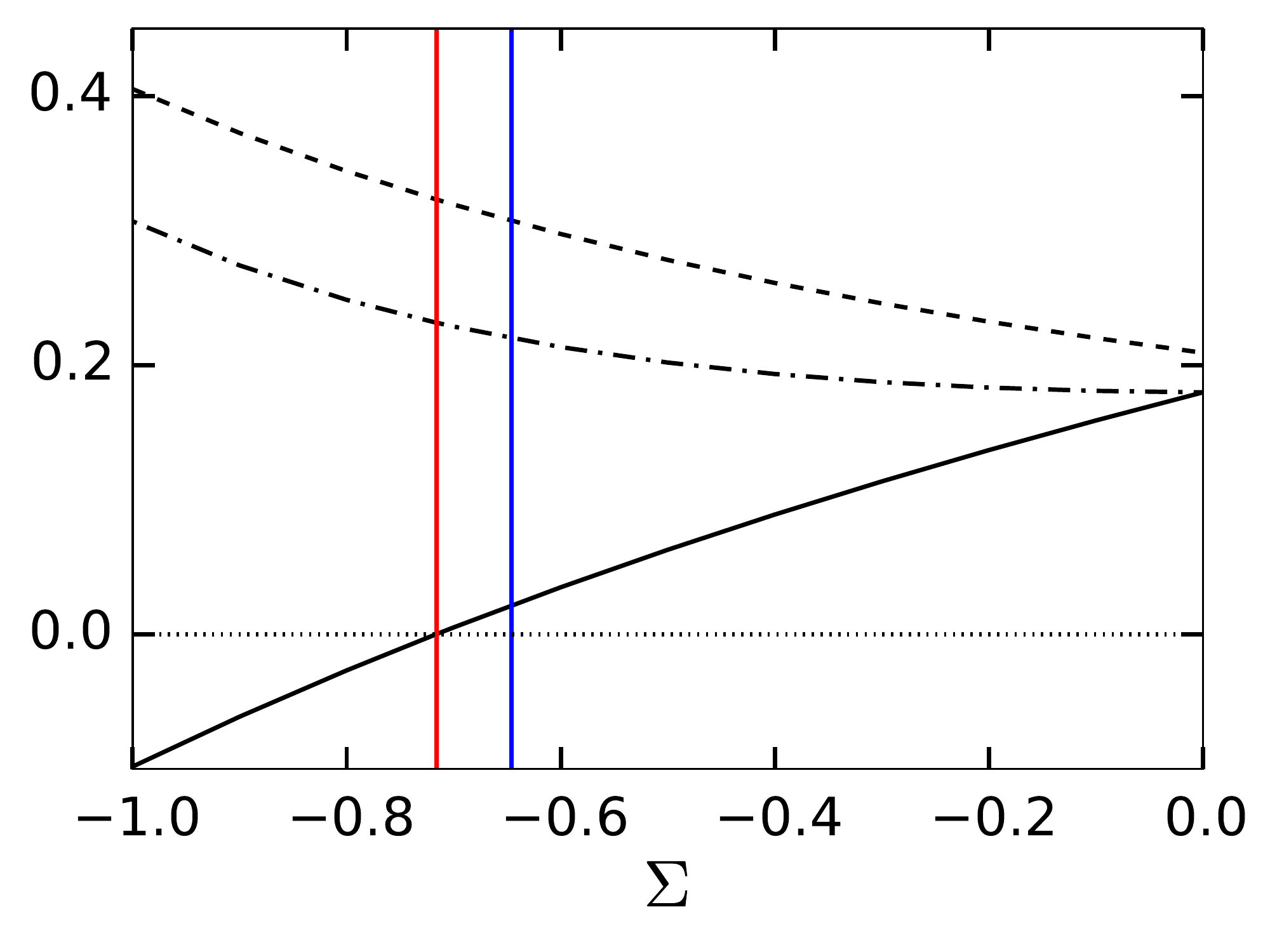}
    \caption{
(Color online) Domain wall energy and excess area. Continuous curve: $\xi _{DW}$. Dashed curve: $\alpha _{DW}$. Dotted-dashed curve: $e _{DW}$. The vertical lines indicate  $\Sigma _c$ (left red), and $\Sigma _{nl}$ (right blue). 
		}
        \label{fig:figure5}
	\end{center}
\end{figure}

\begin{figure}[ht]
    \begin{center}
            \includegraphics[width=0.5\textwidth]{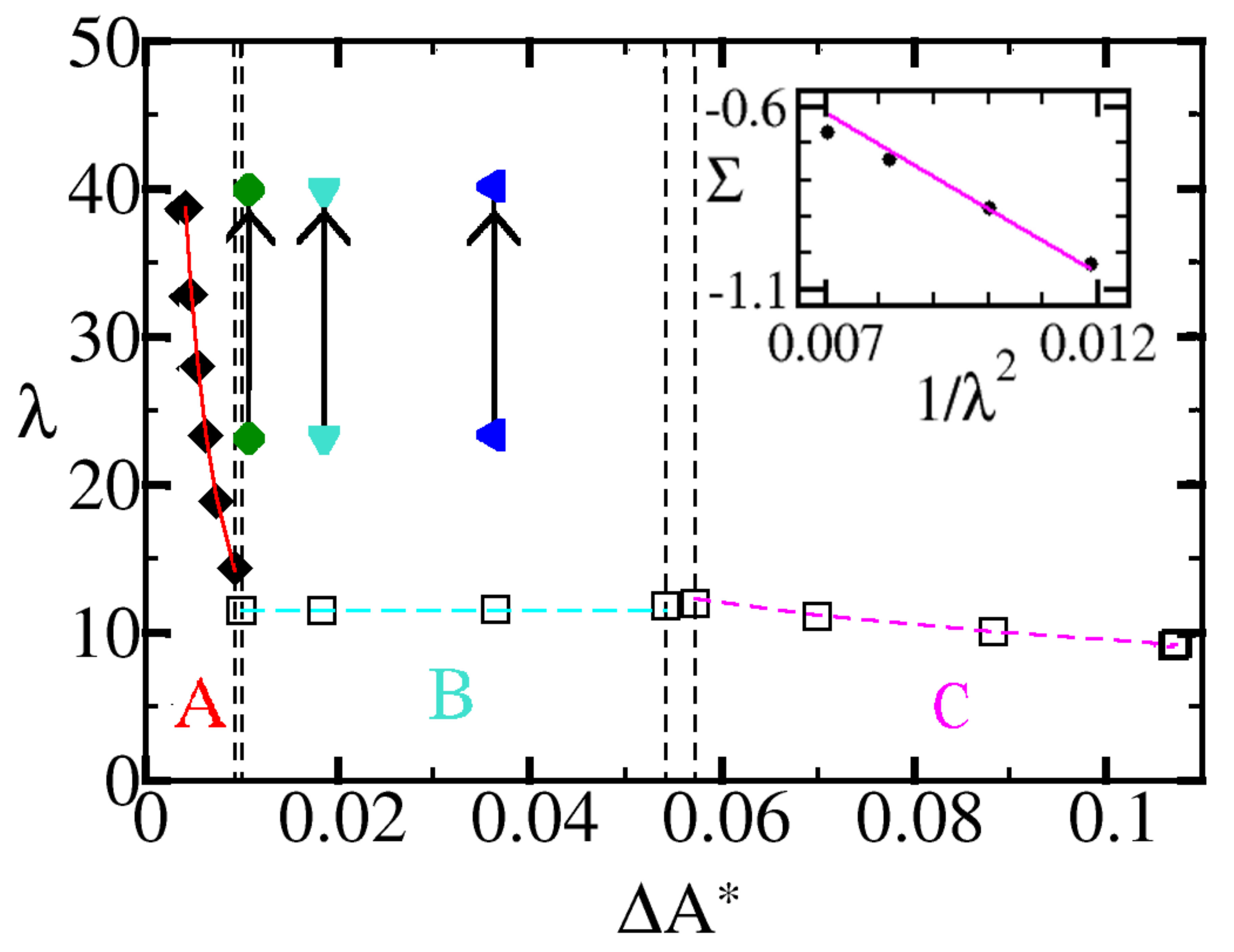}
    \caption{
(Color online) Typical length scale $\lambda$ at long times as a function of $\Delta A^*$ for $L=L_X=L_Y=800$. 
Symbols: $\MyDiamond[draw=black,fill=black]$ for $\lambda _{flat}$,
and $\square$ for $\lambda _{1wr}$. The critical values
of the normalised excess area
$\Delta A^*_c$ and $\Delta A^*_{nl}$ 
are indicated with dashed black lines
(these lines are doubled to indicated the accuracy 
of our measurement of the transition).
Thin red curve and thin magenta curve: analytical results. 
Inset: linear dependence of $\Sigma$ on $1/\lambda^2$. 
Symbols for simulation results \tikzcircle[fill=black]{2pt}. 
Magenta line: $\Sigma=-88/\lambda ^2$.
		}
        \label{fig:figure4}
	\end{center}
\end{figure}

Using  independent 1D simulations of the SH equation 
in a periodic box
with two opposite domain walls, we have determined
$e_{DW}$ as a function of $\Sigma$ in steady-state, as seen in Fig. \ref{fig:figure5}.
In particular, we find $e_{DW}^c\approx 0.2315$. 
Inserting this value in Eq. (\ref{e:bar_lambda_flat}), we obtain
a prediction of $\bar{\lambda}_{flat}$ in good agreement with numerical results, 
as shown by the left red curve on Fig. \ref{fig:figure4}.

We now discuss why the tension converges to the special value $\Sigma_c$.
In order to investigate this point, we make use of an effective model 
for domain wall motion.
The derivations are inspired from that of Ref. \cite{LeGoff2015StatMech}, 
and some details are reported in Appendix \ref{a:motion _by_curvature}. 
As expected, we find that domain wall
motion can be driven either by wall-wall interactions
or by curvature, and the local normal velocity of the wall obeys
\begin{align}
V_n=-\frac{1}{\alpha_{DW}} \Big( [U_0]^+_- + K \xi_{DW} \Big).
\label{e:vn_walls}
\end{align} 
where $K$ is the local domain wall curvature,
and $[U_0]^+_-$ is an interaction term
(see Appendix \ref{a:motion _by_curvature} for its detailed expression).
As discussed above, interactions between two domain
walls are known to be oscillatory \cite{LeGoff2015StatMech},
and to decay
exponentially. They become negligibly small when wall-wall distances 
exceed a few domain wall thicknesses.

In regime A, domain walls are seen to be
far apart, and as a consequence, domain wall motion
should be mainly driven by curvature.
Such curvature-driven domain wall motion
leads to changes in the system configuration
that reduce the total length $L_{DW}$ of the domain walls
when $\xi_{DW}>0$, i.e. when $\Sigma>\Sigma_c$ from Fig. \ref{fig:figure5}.
Indeed, using the geometric relation 
$\partial_tL_{DW}=\int d\ell_{WD}K V_n$,
and inserting the expression of $V_n$ from Eq. (\ref{e:vn_walls})
neglecting the interaction term, we find 
$\partial_tL_{DW}=-(\xi_{DW}/\alpha_{DW})\int d\ell_{WD}K^2<0$.
Since the total excess area $\approx L_{DW}\alpha_{DW}$ is conserved, 
this decrease of $L_{DW}$ leads 
to an increase of the excess area $\alpha_{DW}$ 
stored per unit length in domain walls.
Such a change in $\alpha_{DW}$ is
intuitively associated to a decrease of the tension $\Sigma$
toward more negative values. 
Indeed, we expect $\partial_\Sigma\alpha_{DW}\leq 0$, i.e.,
more positive tensions
correspond to pulling the membrane out from the domain walls which 
decrease the excess area inside the domain wall,
while more negative tensions correspond to pushing  
towards the domain walls leading to an increase of excess area. 
The inequality $\partial_\Sigma\alpha_{DW}\leq 0$ 
is confirmed by 1D simulations of the SH equation
in steady-state reported in Fig. \ref{fig:figure5}.

As a summary, curvature-driven wall motion
leads to an increase of $\alpha_{DW}$ accompanied by 
a decrease of the tension $\Sigma$.
This decrease is governed by the equation 
\begin{align}
\partial_T \Sigma = 
\frac{\xi _{DW} \int d\ell_{DW} K^2 
+ \int d\ell_{DW} [U_0]^+_- K}{L_{DW} \partial_\Sigma \alpha_{DW}}.
\label{e:dSdT}
\end{align} 
where $\ell_{DW}$ is the arclength along the domain walls,
and the integration runs over all domain walls.
This relation is a consequence of area conservation,
and its derivation is reported 
in Appendix \ref{a:motion _by_curvature}.
Once again, we neglect the interaction terms proportional to $[U_0]^+_-$.
As seen in Fig. \ref{fig:figure5},
$\xi_{DW}$ is an increasing function of $\Sigma$.
Since $\partial_T \Sigma$ is proportional to 
$\xi_{DW}$, and recalling that $\partial_\Sigma\alpha_{DW}\leq 0$,
Eq. (\ref{e:dSdT}) shows
that $\Sigma$ will decrease up to the point where $\xi_{DW}=0$,
where it reaches the constant $\Sigma_c$.
Since $\xi_{DW}\rightarrow 0$, the motion
of domain walls freezes from Eq.~(\ref{e:vn_walls}). 

As already mentioned above, this discussion is based on the assumption that
interactions between domain walls are weak.
Since interactions decrease exponentially with the distance,
such an assumption should be valid in the limit 
where the distance between domain walls is large enough.
However, as $\Delta A^*$ increases, this distance
decreases, and interactions between walls should become relevant.
Indeed, a different regime, discussed in the next section and hereafter denoted as regime B
appears for larger $\Delta A^*$. 

\begin{figure}
    \begin{center}
            \includegraphics[width=0.5\textwidth]{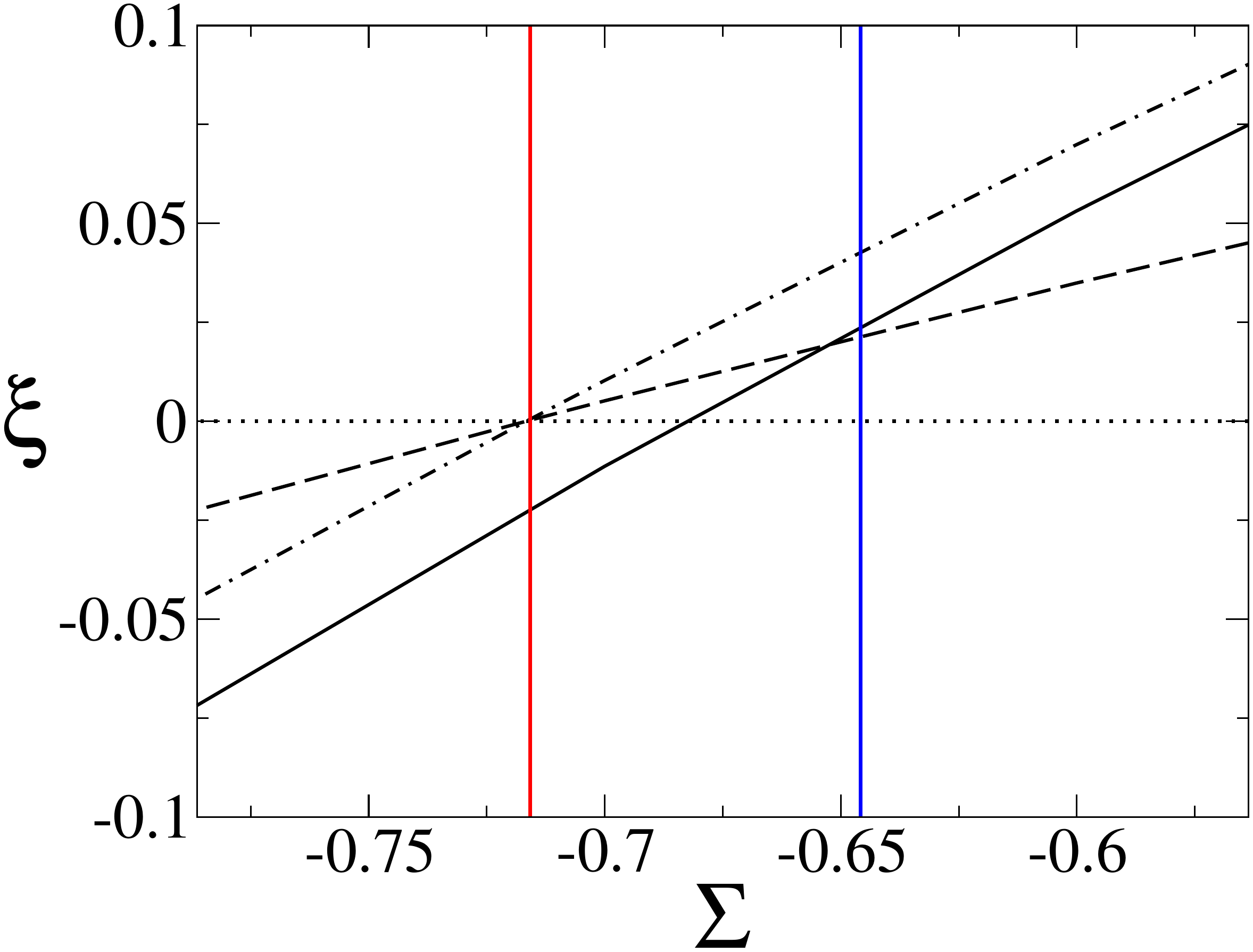}
    \caption{
(Color online) Comparison of the energies
of an isolated wrinkle, and of two domain walls.
Left red vertical line: $\Sigma _c$. 
Right blue vertical line: $\Sigma _{nl}$. 
Solid line: $\xi _{1wr}$. 
Dashed line: $\xi _{DW}$. 
Dashed-dotted line: $2\xi_{DW}$.
		}
        \label{fig:figure6}
	\end{center}
\end{figure}

\section{Intermediate excess area}
\label{s:regimeB}

As announced in the previous sections, a different regime is observed for larger excess
area. This regime, denoted as regime B 
is found for $\Delta A^*_c < \Delta A^* < \Delta A^*_{nl}$,
with
\begin{align}
\Delta A^*_c &=(0.93 \pm 0.03) \times 10^{-2}
\label{e:DeltaA*c_num}
\\
\Delta A^*_{nl} &= (5.53 \pm 0.15) \times 10^{-2}
\label{e:DeltaA*nl_num}
\end{align}
The errors on $\Delta A^*_c$ and $\Delta A^*_{nl}$ are based on the 
difference between the closest upper and lower bounds 
for the transition obtained by long-time simulations
in a system of size $L=L_X=L_Y=800$.

\begin{figure*}
     \centering
     \subfloat[]{\includegraphics[width=0.45\textwidth]{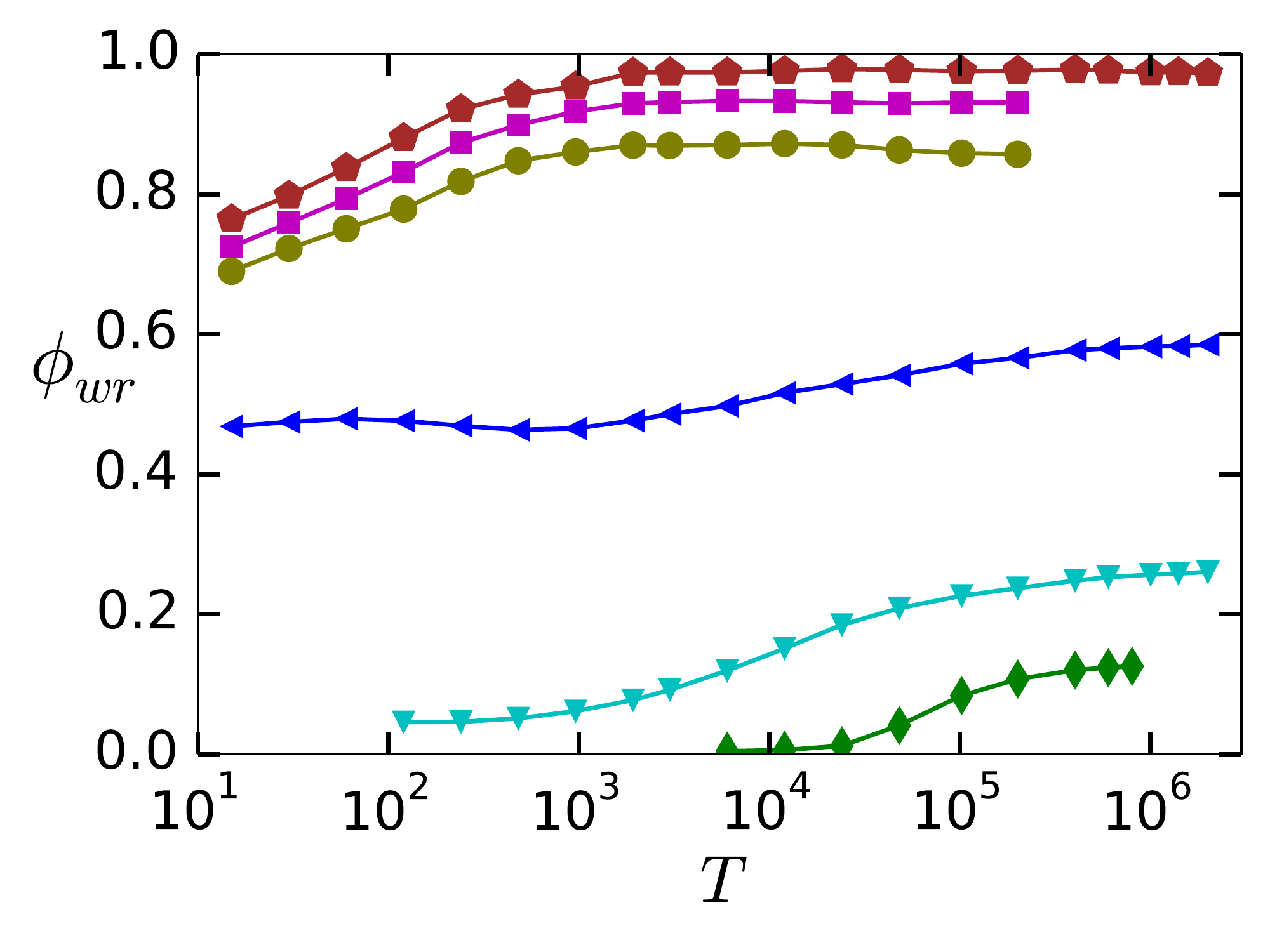}} 
     \subfloat[]{\includegraphics[width=0.45\textwidth]{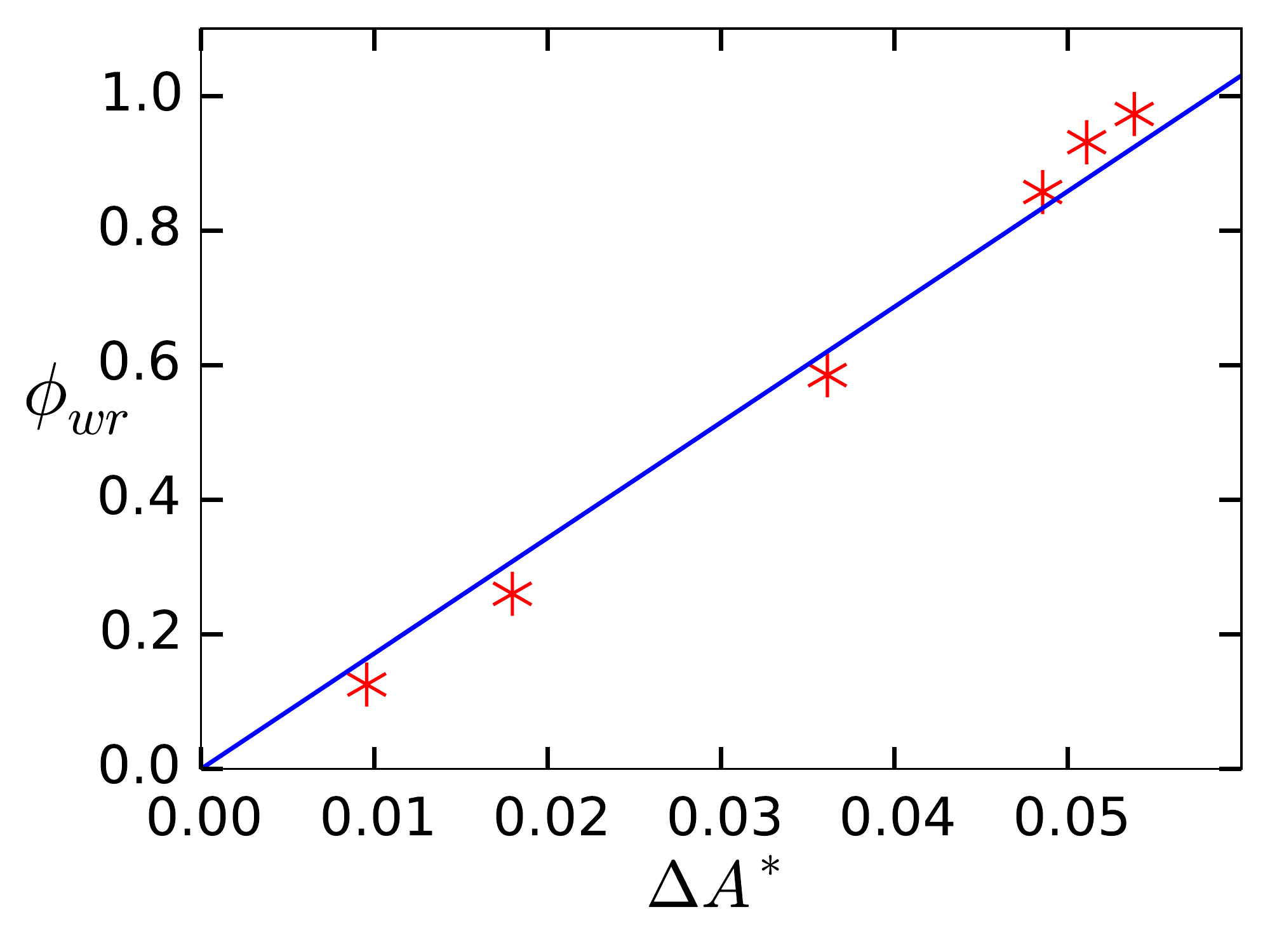}}
     \caption{(Color online) 
     Time-evolution of (a) 
the area fraction covered by wrinkle domains $\phi_{wr}$
     for $L=400$ (top three curves) and $L=800$ (bottom three curves). 
     ({\color{Maroon}\protect\MyPentagon}) $\Delta A^*=5.38 \cdot 10^{-2}$. (\MySquare[magenta,fill=magenta]{2pt}) $\Delta A^*=5.11 \cdot 10^{-2}$. (\tikzcircle[olive,fill=olive]{2pt}) $\Delta A^*=4.85 \cdot 10^{-2}$. ({\color{LightBlue} $\blacktriangleleft$}) $\Delta A^*=3.61 \cdot 10^{-2}$. ({\color{BlueGreen} $\blacktriangledown$}) $\Delta A^*=1.8 \cdot 10^{-2}$. ($\MyDiamond[draw=OliveGreen,fill=OliveGreen]$) $\Delta A^*=0.96 \cdot 10^{-2}$. 
     (b)  Area fraction covered by wrinkle domains at long simulation times. 
     ({\color{red}\Large $*$}) asymptotic value extracted from simulations.
     Blue line: Theoretical prediction from Eq. \eqref{e:phi_w}.
     }
     \label{fig:lphiwr}
\end{figure*}

In  regime B, both flat and wrinkled domains 
coexist and expand perpetually (see regimes B1 and B2 in Figs. \ref{fig:figure2}). 
Numerical results for the dynamics of the tension
and of the size of the domains are shown in Figs. \ref{fig:figure3}(a) and \ref{fig:figure3}(c), 
with the symbols ({\color{red} $\blacktriangle$}),
({\color{orange} $\blacktriangleright$}),
($\MyDiamond[draw=OliveGreen,fill=OliveGreen]$), 
({\color{BlueGreen} $\blacktriangledown$}) and 
({\color{LightBlue} $\blacktriangleleft$}) 
in the order of increasing $\Delta A^*$. 
Regime B corresponds to the symbols 
($\MyDiamond[draw=OliveGreen,fill=OliveGreen]$), 
({\color{BlueGreen} $\blacktriangledown$}) and 
({\color{LightBlue} $\blacktriangleleft$}).
The typical sizes of flat domains $\lambda_{flat}$ and the size of
wrinkle domains $\lambda_{wr}$ both increase with time. 
The method for extracting these lengths from simulation data is
discussed in Appendix \ref{a:lengthscales}.
The growth rates of $\lambda_{flat}$ and $\lambda_{wr}$ 
decrease with increasing $\Delta A^*$. 
However, no universal exponent related to this coarsening process could be found
within our simulations.
In addition, we observe that the fraction $\phi_{wr}$
of the system covered with the wrinkle phase 
reaches a constant at long times, as shown in Fig. \ref{fig:lphiwr}(a).

As seen in Fig. \ref{fig:figure3}(c), the tension in regime B first decreases to a minimum value 
close to or larger than $\Sigma_c$,
and then increases to another constant asymptotic value. 
This value, hereafter denoted as $\Sigma_{nl}$,
corresponds to a point of coexistence
of flat and wrinkled states in our system. 
Following the same procedure as for the definition of 
generalised thermodynamic potentials,
coexistence between two states correspond to the equality 
of Legendre-transformed energy $\Xi$
with respect to the Lagrange multiplier $\Sigma$ conjugate to the 
fixed quantity $\Delta A$.
Coexistence therefore corresponds to the point where the $\Xi$-energy densities 
of the flat and wrinkled states are equal: $\Xi_{FD}/A_{FD}=\Xi_{wr}/A_{wr}$,
where the indexes "FD" and "wr" respectively 
correspond to the flat domains regions
and wrinkles regions.

Note that the energy $\Xi_{FD}$ a priori contains not only the 
contribution of flat domains with $H=\pm H_m$, but also that of domain
walls inside this region. However, the density of domain
walls is low, and we shall only consider the 
contribution of flat domains, for which $\Xi_{FD}=0$
(such a cancellation relies on the assumption that the 
minimum $H_m$ is far enough form the walls for the short-range
potential $U_d$ to be negligible at $H=H_m$, i.e., $1-H_m\ll d$).
Following the same lines,
the energy within the wrinkle phase contains a contribution due to wrinkle bending and wrinkle defects, which is also neglected.
We therefore end up with a simpler
condition of coexistence $\Xi_{wr}=0$, calculated for a periodic phase of parallel rolls.
This is exactly the condition of nonlinear stability
of domain walls with respect to the formation of a wrinkle phase in SH
as discussed in Ref. \cite{Hagberg2006}, leading to $\Sigma_{nl}\approx -0.9225 H_m$ 
in agreement with our simulation results.

Using this concept of coexistence, together with 
the observation of the variation of the tension with time, 
we propose a scenario in three stages
for regime B. First, the dynamics looks like that of regime A, and
the evolution is dominated by domain wall motion, leading to a decrease of $\Sigma$
toward $\Sigma_c<\Sigma_{nl}$.
Second, once $\Sigma<\Sigma_{nl}$, the flat states become unstable with respect to the formation
of the wrinkle state. In simulations at low excess area, i.e., close to regime A,
this instability usually occurs as follows:
two domain walls collide and form an isolated wrinkle,
which grows by zipping more domain walls.
This is consistent with the inequality 
$\xi_{1wr}<2\xi_{DW}$, where $\xi_{1wr}$ is the $\Xi$-energy per unit length of 
a straight and isolated wrinkle, which can be checked on Fig. \ref{fig:figure6}.
For larger excess area, the separation between the two initial stages
is less clear since many domain walls are already close to each other initially.
Finally, in the third stage, coarsening occurs with coexistence of 
wrinkle domains and flat domains.
It is tempting to speculate that the coarsening process
is driven by the decrease of the length of frontiers between the coexisting wrinkled state
and the flat state. However, annihilation of defects within the wrinkled state
and motion of simple domain walls between flat domains
could also play a role.

When the excess area is increased, the fraction of the system covered
by wrinkles increases. When the excess area exceeds a threshold value,
the full system is covered by wrinkles, 
and a different regime, denoted as regime C
is found.



Some analysis of the fraction $\phi_{wr}$ of the system
occupied by wrinkles in regime B is possible assuming 
once again that the contribution
of defects such as domain walls in the flat phase and
defects in the wrinkle phase are negligible.
The wrinkle state
is then composed of parallel rolls of  
wavelength $\lambda_{1roll}^{nl}$ where the superscript $nl$ indicates
that this quantity is evaluated for $\Sigma=\Sigma_{nl}$. 
We define wrinkle length $L_{wr}$, i.e., the 
total wrinkle length summed over all wrinkles
(formally, this can be defined as, e.g., the total length of all the lines of local maximums
of the wrinkles in the whole system).
Assuming that  all the excess area is stored in the 
wrinkle phase, one has $\Delta A=\alpha_{1roll}^{nl}L_{wr}$.
The area covered by wrinkles
then reads $A_{wr}=L_{wr}\lambda_{1roll}^{nl}$. 
Combining these relations, we find that
the fraction $\phi_{wr}=A_{wr}/A_{syst}$ of the system covered by the wrinkle state
is proportional to the normalised excess area $\Delta A^*$:
\begin{align}
\phi_{wr}
=\Delta A^*\frac{\lambda_{1roll}^{nl}}{\alpha_{1roll}^{nl}}.
\label{e:phi_w}
\end{align}
In order to determine the quantities appearing in the 
right hand side of Eq. \eqref{e:phi_w}, we simulated 
the SH equation with $\Sigma=\Sigma_{nl}$ in 1D
in a box of size $L=800$. 
We found that a stable wrinkle profile is reached 
with 66 or 67 wavelengths. 
Taking the average of these values, we obtain 
 $\lambda_{1roll}^{nl}=12.03\pm0.09$, 
$\alpha_{1roll}^{nl}=0.7006 \pm 0.0042$,
and $e^{nl}_{1roll}=0.4575 \pm 0.0025$. 
These numerical results in 1D confirm that $\Xi_{wr}$ 
in the roll phase vanishes at $\Sigma=\Sigma_{nl}$.
Indeed, 
$\xi^{nl}_{1roll}=e^{nl}_{1roll}+\Sigma_{nl}\alpha_{1roll}^{nl}\approx (5.1 \pm 0.2) \cdot 10^{-3}$.
In addition, we find $\lambda_{1roll}^{nl}/\alpha_{1roll}^{nl}=17.17 \pm 0.23$.
Using this result in the prediction Eq. \eqref{e:phi_w} provides good agreement 
with simulations, as seen in Fig.~\ref{fig:lphiwr}(b).

Assuming that the transition to the frozen wrinkle state 
at large excess area corresponds to the filling of the system
with the wrinkle state $\phi_{wr}=1$, we obtain a prediction
for the critical excess area corresponding to the transition
to regime C:
\begin{align}
\Delta A^*_{nl}=\frac{\alpha_{1roll}^{nl}}{\lambda_{1roll}^{nl}}.
\end{align}
Using the numerical values reported above,
we find $\Delta A^*_{nl}=\alpha_{1roll}^{nl}/\lambda_{1roll}^{nl}=(5.83\pm 0.78) \cdot 10^{-2}$ which is very close to the value of $\Delta A^*_{nl}$ observed in the simulations, Eq. \eqref{e:DeltaA*nl_num}.



\section{Large excess area}
\label{s:regimeC}

Regime C 
is obtained for $\Delta A^* > \Delta A^*_{nl}$. 
In this regime, after some transient dynamics, 
the adhesion pattern is frozen into labyrinths  
as shown in Fig. \ref{fig:figure2}C.
Such labyrinthine patterns have also been observed in the 
case of the SH equation \cite{LeBerre2002}.

In our simulations of regime C, the tension again converges to a stationary value.
However, as opposed to the previous cases in regimes A and B,
the asymptotic tension depends on $\Delta A^*$.
Furthermore, we observe in the simulations that, 
as the excess area $\Delta A^*$ increases in regime C, 
the width of the wrinkles $\lambda=\lambda_{1roll}$
decreases, whereas the maximum height  
and the variance of the membrane profile $\langle H^2 \rangle$ increase.

In order to analyze this behavior, we perform the changes of variables $\chi=\zeta/\lambda_{1roll}$, and 
$\eta = H/\langle H^2 \rangle ^{1/2}$. The excess area is then written as
\begin{align}
\Delta A^*=\frac{\langle H^2 \rangle}{2\lambda_{1roll} ^2} \int _0^1 d\chi (\partial _\chi \eta)^2
\label{e:scaled_DeltaA*_regimeC}
\end{align}
where the integral in the right hand side only depends on the shape
of the wrinkle profile. 
An inspection of Eq. \eqref{e:scaled_DeltaA*_regimeC} shows that, 
if the shape of the wrinkle does not vary much,
the quantity $\Delta A^{* 1/2} \lambda_{1roll} / \langle H^2 \rangle^{1/2}$ should be a constant.
Indeed, the relation $\Delta A^{* 1/2} \lambda_{1roll} / \langle H^2 \rangle^{1/2} \approx 3.6$
is in good agreement with numerical results. Using the value
of $\langle H^2\rangle^{1/2}$ measured in simulations,
this expression provides the dashed magenta line on the main plot of Fig.~\ref{fig:figure4}.

Assuming that the profile of the wrinkles is sinusoidal $H\sim\sin(2\pi\zeta/\lambda_{1roll})$,
we find $\eta=2^{1/2}\sin(2\pi\chi)$, and $\int _0^1 d\chi (\partial _\chi \eta)^2=4\pi^2$,
suggesting that 
\begin{align}
\frac{\Delta A^{*\,1/2} \lambda_{1roll} }{ \langle H^2 \rangle^{1/2}}= 2^{1/2}\pi\approx 4.4 .
\label{eq:wrinkles_DeltaA*_lambda_h2}
\end{align}
The sine ansatz therefore leads to an overestimation 
of the constant, showing that the profile of the rolls is different 
from a simple sinusoidal profile.



Furthermore, the asymptotic value of the tension is controlled by the 
balance between the negative tension $\Sigma$ which tends to store excess area
by forming the wrinkles,
and the bending rigidity which tends to flatten the membrane.
A simple balance between the two terms in the energy Eq. (\ref{e:Xi_energy_rescaled}) 
suggests that $\Sigma H^2/\lambda_{1roll}^2\sim H^2/\lambda_{1roll}^4$, 
or  $\Sigma \sim 1/\lambda_{1roll}^2$.
Simulation results reported in the inset of Fig. \ref{fig:figure4}
indicate that this relation is in fair agreement
with the numerical results with $\Sigma \lambda_{1roll}^2 \approx -88$.

If we assume again a sine profile for the membrane rolls,
the potential energy density is not affected by the wavelength.
The contributions that depend on the wavelength $\lambda_{1roll}$ are the bending
energy and the tension contribution. Minimizing the 
energy with respect to the wavelength, we find
\begin{align}
\Sigma \lambda_{1roll}^2 =-8\pi^2\approx -79.
\label{eq:wrinkles_Sigma_lambda2}
\end{align}
This relation between the tension and the wavelength
appears in better agreement with the simulations as
compared to Eq. (\ref{eq:wrinkles_DeltaA*_lambda_h2}),
but is still not very accurate.

In the following, we proceed further with the sine ansatz
to predict the wavelength, amplitude and tension
from a direct minimization of the energy density.
The details of the derivations are reported in Appendix~\ref{a:sine_ansatz_regimeC}.
Depending on the excess area, we find two regimes. 
In both regimes Eqs. (\ref{eq:wrinkles_DeltaA*_lambda_h2},\ref{eq:wrinkles_Sigma_lambda2}) are valid.
First, when the amplitude of the sine
profile is small enough, the amplitude and the wavelength 
are both varied to minimise the energy density. However,
for large enough excess area, the extrema of the sine profiles 
touch the walls. In this case, the short range repulsion potential $U_d$
prevents the membrane from crossing the wall. We therefore fix the amplitude
and minimise the wavelength only. The crossover to this wall-contact regime 
occurs for $\Delta A^*=\Delta A^*_{wc}$ with
\begin{align}
\Delta A^*_{wc}=\frac{1}{8} (3-4H_m^2)^{1/2}.
\label{eq:Delta_A*_st}
\end{align}
The details of the calculations are reported in Appendix~\ref{a:sine_ansatz_regimeC}.

\begin{figure}
     \includegraphics[width=0.4\textwidth]{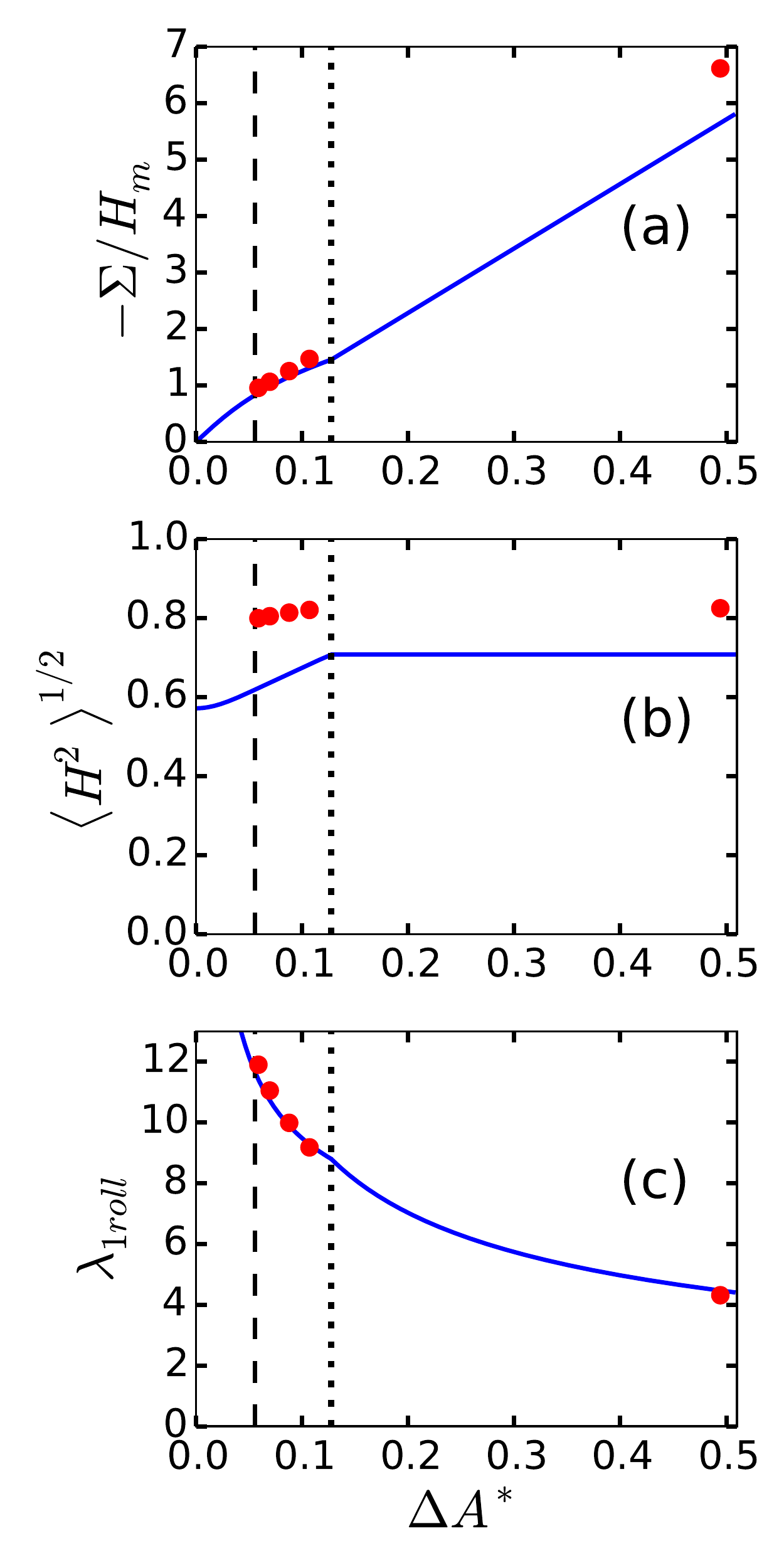}
     \caption{(Color online) 
     (a) Tension, (b) amplitude and (c) wavelength of the wrinkles in regime C.
     Red dots represent the simulation results.
     The solid blue lines are the result of the sine-profile ansatz. The dashed
     and dotted vertical lines respectively indicate the start of regime C at 
     $\Delta A^*= \Delta A^*_{nl}$, and the crossover to the wall-contact regime
     at $\Delta A^*= \Delta A^*_{wc}$.
     }
     \label{fig:sine_ansatz_regimeC}
\end{figure}

The results of the sine-profile ansatz, shown on Fig. \ref{fig:sine_ansatz_regimeC},
reproduce the trends obtained with the full simulations in regime C. 

Note that the above value of $\Delta A^*_{wc}$ corresponds to a crossover
rather than a sharp transition in the simulations. In addition, our convenient
decomposition of the potential into a smooth double-well
potential and a sharp short-range repulsion is valid only  when the minimum
of the potential is far enough from the walls. In contrast, when $H_m$
is close to $1$, the short
range repulsion $U_d$ affects the membrane profile in the potential well.
One consequence of this is the inconsistency of Eq.~\eqref{eq:Delta_A*_st}
when $H_m>3^{1/2}/2\approx 0.866$. Our choice
$H_m=0.7$ in simulations however corresponds to the regime where Eq.~\eqref{eq:Delta_A*_st}
is valid.

\section{Impermeable walls}
\label{s:impermeablewalls}

\begin{figure*}
    \begin{center}
            \includegraphics[width=0.6\textwidth]{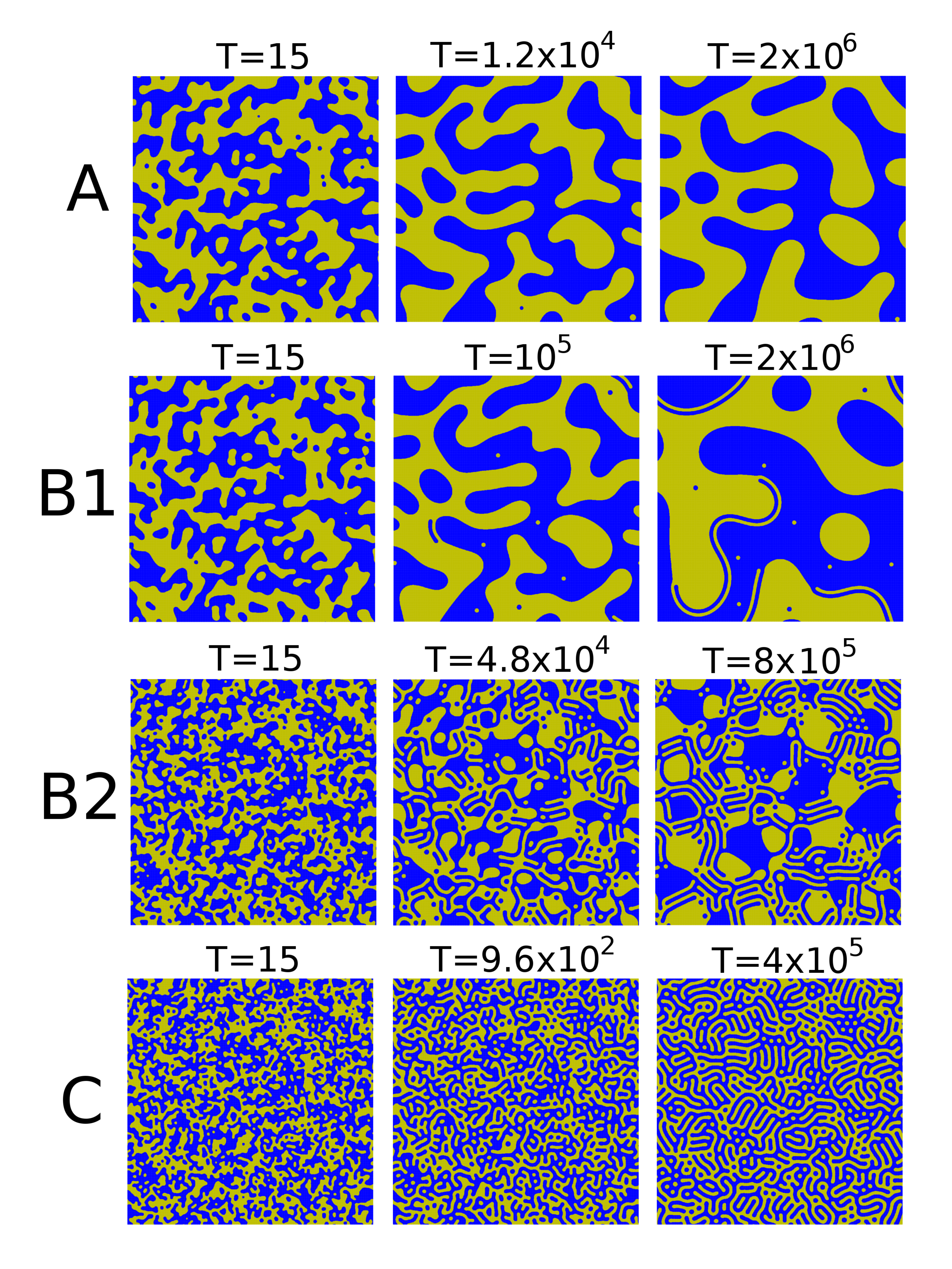}
    \caption{
(Color online) Membrane adhesion dynamics for impermeable wall for various excess areas. Yellow: adhesion patches
 on the upper wall. Dark blue: adhesion patches on the lower wall. Regime A with freezing
  of flat domains for small excess area : $\Delta A^*=0.74 \times 10^{-2}$.  
Regime B with coexistence of the flat-domain phase and wrinkle phase
 with coarsening for intermediate excess area: B1 with $\Delta A^*=0.88 \times 10^{-2}$,
 and B2 with $\Delta A^*=3.61 \times 10^{-2}$. 
 Regime C with frozen wrinkles for larger excess area $\Delta A^*=5.68 \times 10^{-2}$. 
 System size $L \times L=400\times400$.	}
        \label{fig:conserved}
	\end{center}
\end{figure*}

In this section, we discuss the limit of 
very impermeable walls with $\bar\nu\ll 1$.
The full lubrication equations are reported in Appendix \ref{a:lubrication}.
These equations include not only a term accounting for the conservation
of the total flow,
as already found in a one-dimensional model \cite{LeGoff2014}, 
but also a space-dependent tension that drives tangential
forces along the membrane due to local area conservation.
For the sake of simplicity, we neglect these two terms,
and discuss a simplified model which is a straightforward
transposition of Eq. (\ref{eq:hpermeable},\ref{eq:sigmapermeable}) to the case of conserved dynamics:
\begin{subequations}
\begin{align}
\partial _t h = \nabla \Big({\cal M}(h) \nabla (\kappa \Delta ^2 h - \sigma_0 \Delta h + {\cal U}^\prime(h))\Big),
\label{eq:dhdt_impermeable}
\end{align}
\begin{align}
\sigma_0 = \frac{\int \int dxdy \nabla \Big({\cal M}(h) \nabla (\kappa \Delta ^2 h + {\cal U} ^\prime (h))\Big)\Delta h}{\int \int dx dy \nabla ({\cal M}(h) \nabla \Delta h)\Delta h}.
\label{eq:sigmaimpermeable}
\end{align}
\label{eq:impermeable}
\end{subequations}
The nonlinear mobility \cite{LeGoff2014}
\begin{align}
{\cal M}(h)=\frac{h_0^3}{24\mu} \left[ 1-\frac{h^2}{h_0^2} \right]^3 
\end{align}
expresses the slowing down of the dynamics when the membrane approaches the wall at $h=\pm h_0$.
This slowing down is caused by the increase of viscous dissipation when squeezing a thin 
liquid film separating the membrane from the wall. 

The simulations are performed in rescaled units that are defined
in the same way as for the permeable limit, except for the rescaling
of time $T={\cal U}_0^{3/2}t/(24\mu\kappa^{1/2})$. In addition,
we defined a rescaled mobility $M(H)=(1-H^2)^3$ via the relation ${\cal M}(h)=M(H)(h_0^3/(24\mu))$.

We have performed simulations both with $M(H)=(1-H^2)^3$, and in the simplified case $M(H)=1$.
Simulations performed with $L=400$ and with $M(H)=1$ 
exhibit a behavior similar to 
that obtained with $M(H)=(1-H^2)^3$ 
but are computationally ten times faster. This observation
is consistent with previous simulations 
with the 1D model \cite{LeGoff2015PRE}. 
Thus, in order to minimise finite size 
and finite time effects,
we have performed a systematic study with $M(H)=1$ for 
a system size $L=800$. 

The evolutions of $\lambda_{flat}$, $\lambda_{wr}$ 
and $\Sigma$ are shown in Figs. \ref{fig:figure3}(b) and \ref{fig:figure3}(d) respectively. 
Due to the slow dynamics the results
are less conclusive than the high permeability case. 
However, the simulations exhibit similar trends as 
in the large permeability limit, and
we recover the three regimes A, B and C discussed in the previous sections. 
The main differences are that (i) the dynamics 
is much slower, and (ii) 
we obtain a smaller value 
for the transition to coarsening $\Delta A^*_c = (0.4 \pm 0.03) \times 10^{-2}$ 
as compared to the large wall permeability limit.

Furthermore, as in the case of permeable walls, we find that
tensionless dynamics, obtained by solving 
Eq. (\ref{eq:sigmaimpermeable}) with $\Sigma=0$
leads to the same coarsening process as the Cahn-Hilliard
equation, with the domain size growing as $t^{1/3}$.
The results are reported in Fig.\ref{fig:tdgl4ch4cm}.

As a summary, the conserved dynamics discussed in this section
using a simplification of the non-conserved case
behaves in a way which is similar to the non-conserved case.




\section{Discussion}
\label{s:discussion}

\subsection{Finite size and initial conditions effects}
\label{s:finite_size_effects}

Since we do not have access to infinitely long
times and infinitely large system sizes in simulations, 
we cannot make a final statement
regarding the fact that frozen states
do evolve slowly or are absolutely frozen at very long times.
This is particularly important for the evaluation
of $\Delta A_c$. Indeed, our observations show that the 
transition from regime A (frozen) to regime B (coarsening) 
can be started via the formation of a single wrinkle resulting
from the collision between two domain walls
somewhere in the system. Thus, if the probability for such
an event to occur per unit area is small but finite, 
then the threshold  $\Delta A^*_c$ should decrease
to zero as the system size increases to infinity.

To check the possible finite-size effects on $\Delta A^*_c$, 
we ran simulations for different system sizes 
$L=400$ and $L=800$ for small $\Delta A^*$. 
In both cases, we obtained the same values  Eq. \eqref{e:DeltaA*c_num} 
for $\Delta A^*_c$ in large permeability limit. 
However, the case of impermeable walls seem to suffer
stronger finite size effects. We indeed find a smaller
value for the largest system
$\Delta A^*_c = (0.4 \pm 0.03) \times 10^{-2}$ for $L=800$,
as compared to $\Delta A^*_c = (0.81 \pm 0.07) \times 10^{-2}$ for $L=400$. 
Hence, our simulation sizes
do not permit to reach a definitive 
conclusion about the existence of a finite
limit for $\Delta A^*_c$ for infinitely large systems.

In addition, the probability to
form wrinkles and to trigger the transition from the frozen regime A to
regime B could be influenced by initial conditions.
We have checked the sensitivity of the $A-B$ transition for $L=400$,
using different random initial conditions described
in Appendix \ref{a:initial_conditions}.
We find a threshold which is similar, but slightly different 
$\Delta A_c^* = (0.835\pm0.035)\times 10^{-2}$ 
for the nonconserved model Eqs. (\ref{eq:hpermeable_norm}, \ref{eq:sigmapermeable_norm}) 
and 
$\Delta A_c^* = (0.77\pm0.03)\times 10^{-2}$ 
for the conserved model with constant mobility Eqs. (\ref{eq:dhdt_impermeable}, \ref{eq:sigmaimpermeable}) with $M(H)=1$.

In summary, our simulations do not allow us to 
reach a conclusion with respect to the existence
of regime A in an infinitely large system. It is however tempting
to speculate that, based on the observation that the wrinkle
phase can form as soon as a single wrinkle appears,
regime A should lead to 
regime B for infinitely large systems
for arbitrary low values of $\Delta A^*$. 
Note that any physical experiment will also be controlled
by finite size effects and finite time observations.
As a consequence, our simulations suggest 
that the frozen phase at low excess area (regime A)
should be observable at least in finite size systems.

In contrast, our interpretation of the transition from regime B
to regime C at high excess area
is based on the filling of the system by the wrinkle phase.
Such a transition is not triggered by an isolated event
as in the transition to coarsening at low excess areas,
and better self-averaging is expected, leading to 
smaller finite size effects, and little sensitivity to intial conditions.

To investigate finite-size effect on $\Delta A^*_{nl}$, 
we ran simulations for systems of different sizes $L=200$ and $L=400$ 
for large $\Delta A^*$. We obtain the same values 
$\Delta A^*_{nl}$ (given in Eq. \eqref{e:DeltaA*nl_num}) 
for both permeability limits. These results rule out the possibility
of an influence of finite-size effect on the transition
from regime B to regime C.

\subsection{Labyrinthine pattern vs parallel rolls in the wrinkle phase}
\label{s:finite_size_effects}

In simulations, the wrinkle phase, which appears
in regimes B and C seems to be more disordered as the excess area
is increased. Indeed, for very small excess areas in regime B, 
when all wrinkles are formed from a single initially isolated
wrinkle, the wrinkle phase looks similar to
a roll phase with a low density of defects.
However, when the excess area increases,
more defects appear in the wrinkle phase.
Finally, in regime C the wrinkle phase is composed of a
disordered labyrinthine pattern.


In simulations of the SH equation by Le Berre \etal \cite{LeBerre2002}, 
labyrinthine patterns were found for $\Sigma_0 < \Sigma < \Sigma_c$ 
where $\Sigma_0= -2.8284H_m$. For $\Sigma < \Sigma_0$, 
these authors only found 
parallel-rolls. 
Moreover, Le Berre \etal \cite{LeBerre2002} show that the
roll pattern is always more stable than the labyrinthine pattern,
i.e. rolls have  lower energy. Hence, the labyrinthine pattern
can be seen as a metastable 
state in which the system can be trapped.

In regime B, we  have $\Sigma>\Sigma_0$ at all times, indicating
that the system can always be trapped in a meta-stable
disordered state.
Our results suggest that, although rolls are more stable,
the order in the wrinkle phase is controlled
by the  domain-size on which the wrinkle phase is forming in regime B. Since
this domain size decreases with increasing $\Delta A^*$,
the related correlation length decreases,
and disorder increases for increasing $\Delta A^*$
in the wrinkle phase of regime B.

In regime C, short-range disorder arises from the
initial formation of microscopic domains, leading
to a labyrinthine pattern.
Since $\Sigma$ in steady-state
decreases with increasing excess area $\Delta A^*$ in regime C,
we tried to increase $\Delta A^*$ to reach the roll phase.
We therefore simulated  a membrane with large excess area $\Delta A^* = 0.4943$ 
in the limit of large permeability. 
In this case, the steady-state tension is found to be $\Sigma = -4.63 < \Sigma_0$. 
However, the system still formed labyrinthine patterns 
and no parallel-rolls in our system.
Note that the short-range repulsion
$U_d$ could play an important role when the 
membrane area is large. Indeed, the membrane
is then in contact with the wall as discussed in Sec. \ref{s:regimeC} 
and the contribution
of $U_d$ should lead to
deviations from the analogy to the simple SH equation.



\subsection{Conservation of the number of adhesion domains}
\label{s:consevration_domains}

\begin{figure*}[ht]
    \begin{center}
            \subfloat[]{\includegraphics[width=0.48\textwidth]{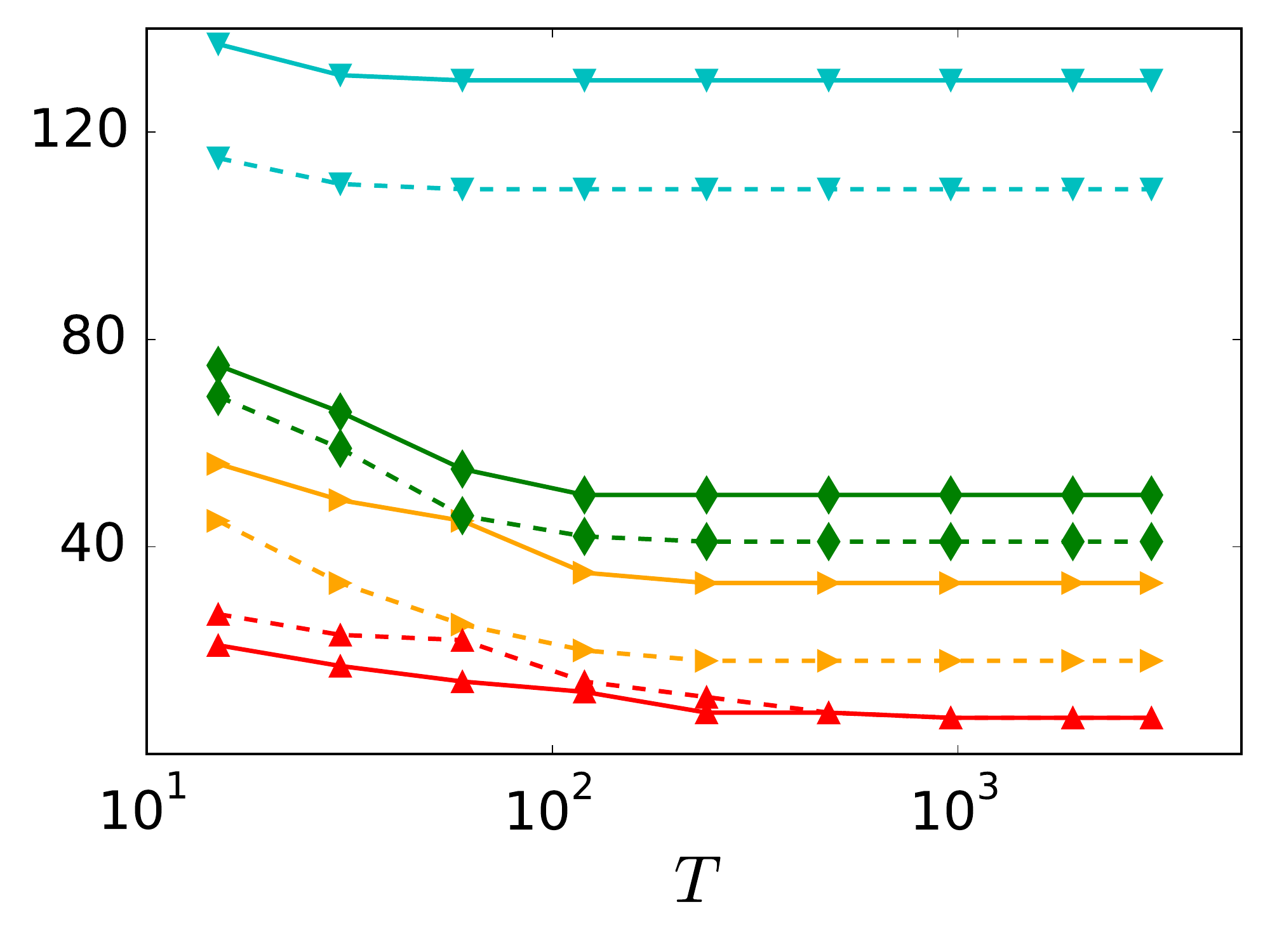}}
            \subfloat[]{\includegraphics[width=0.48\textwidth]{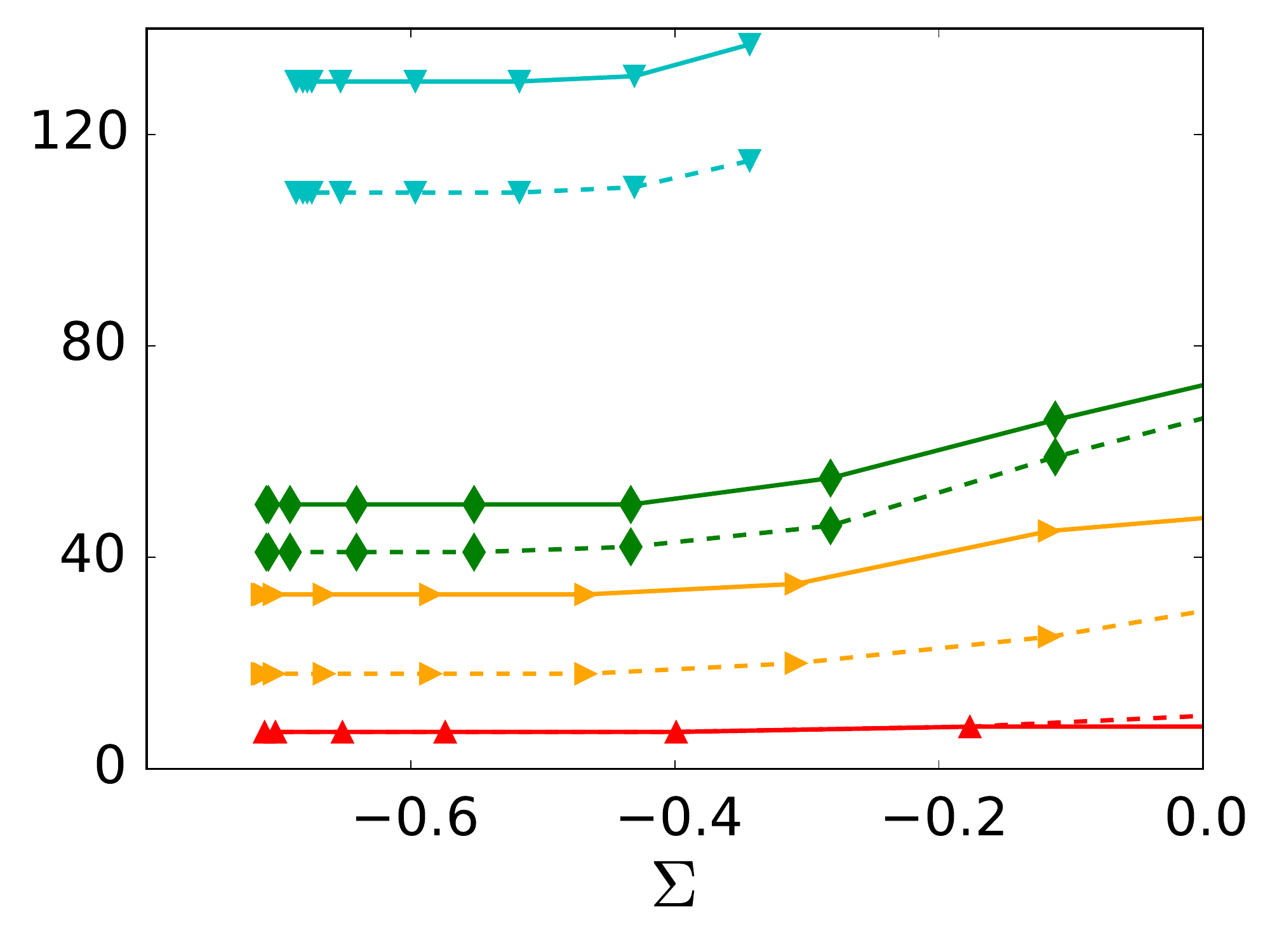}}
    \caption{
(Color online) Number of positive domains (continuous line) and negative domains (dashed line) as a function of (a) time $T$ and (b) membrane tension $\Sigma$ for $L=800$. ({\color{red} $\blacktriangle$}) $\Delta A^*=0.37 \cdot 10^{-2}$. ({\color{orange} $\blacktriangleright$}) $\Delta A^*=0.88 \cdot 10^{-2}$. ($\MyDiamond[draw=OliveGreen,fill=OliveGreen]$) $\Delta A^*=1.07 \cdot 10^{-2}$. ({\color{BlueGreen} $\blacktriangledown$}) $\Delta A^*=1.8 \cdot 10^{-2}$.
}
\label{fig:figure8}
	\end{center}
\end{figure*}

For all three regimes A, B and C, the number of domains decreases initially but is constant at long time. 
The time to reach the constant number of domains decreases as $\Delta A^*$ increases,
as seen from Fig. \ref{fig:figure8}(a).

The origin of this conservation can be traced back to the fact that the disappearance of domains
by shrinking is stopped at small scales due to the formation of  very stable
localised structures. 
These localised structures are so stable that they are sometimes dragged 
on large distances by domain walls or wrinkles without loosing their integrity.
Such stable localised structures have also
been observed in the solution of the SH equation for parameters
corresponding to the range of negative tensions relevant to our
system \cite{Ouchi1996}.

A detailed analysis of the localised structures
is beyond the scope of our study.
However, since the global evolution influences local dynamics via
the tension $\Sigma$, it is tempting to 
propose a conditions under which the conservation 
of the number of domains should mainly be observed:
(i) the domains are already well formed, i.e., there is no roughness
or perturbation smaller than the domain wall width, 
and (ii) the tension is below some critical tension.
From the plot of the number of positive and negative domains
as a function of tension in Fig. \ref{fig:figure8}(b), 
we find that the criterion $\Sigma<-0.552\approx -0.79 H_m$
provides a condition under which the number of domains is always conserved
in our simulations.
The slow dynamics emerging after a short initial relaxation
of the system, and which occurs with tensions around  $\Sigma_c$ 
and $\Sigma_{nl}$ that are lower than this critical tension,
corresponding to a regime where the number of domains is preserved.


\section{Conclusion}


In conclusion, we have developed a lubrication
approach to study the dynamics of 
inextensible membranes confined between two flat attractive walls.
We find that dynamics exhibit three types of regimes
depending on the membrane excess area.
For low excess area (regime A), the membrane
freezes in a configuration with large adhesion
patches on both walls. For intermediate area (regime B),
the membrane exhibits coarsening, with a coexistence
of flat adhesion domains with a wrinkle phase. 
For larger excess area, the membrane freezes into
a labyrinthine wrinkle phase.

We hope that our results can provide hints for the understanding
of the influence of confinement on the dynamics of model lipid membranes.
On a more theoretical level, the model presented
here defines a novel universality class
for phase separation in two dimensions.

In order to gain further insight on the dynamics of 
confined membranes, the role of thermal fluctuations should
be investigated. In one-dimensional models,
such fluctuations were able to restore
the coarsening by allowing the system to 
pass over the energy barriers which were trapping 
the system into metastable frozen states \cite{LeGoff2015PRE}. 
In addition, a study of the dynamics of the conserved model
beyond the simplifications presented above is in order.

Finally, additional ingredients inspired
from biological adhesion, 
such as the mobility and clustering of ligands and receptors~\cite{Fenz2011,Sackmann2014},
or the active remodeling of the cytoskeleton\cite{Pollard2009,Bun2018},
could exhibit non-trivial coupling to the dynamics of the membrane presented here. 




   
\section*{Acknowledgments}
We acknowledge support from Biolub Grant No. ANR-12-BS04-0008. We thank Dr A. K. Tripathi for helpful discussions.



\begin{appendix}

\section{Lubrication limit for a membrane with area conservation}
\label{a:lubrication}

In this section we outline the derivation of the equations which describe the 
dynamics of confined membranes in the lubrication limit.
The derivation is analogous to the one-dimensional model 
by Le Goff \etal \cite{LeGoff2014}. The two novel ingredients are:
(i) the membrane is now two-dimensional, and (ii) we now enforce
membrane area conservation. 

\subsection{General derivation}

We apply the lubrication limit and the small-slope approximation to derive the dynamical equations of the membrane using the standard lubrication expansion. 
This expansion has been used in many studies of thin film dynamics,
and generic details of the calculations can be found in Ref.\nocite{Oron1997}\citenum{Oron1997}.
We assume a separation of scales
\be
x \sim {\cal O} (\epsilon^{-1}), \quad
y \sim {\cal O} (\epsilon^{-1}), \quad
h \sim {\cal O} (1), \quad,
\ee
where $\epsilon$ is a small parameter ($\epsilon\ll1$).
As a consequence, slopes
are small $\partial_xh\sim\partial_yh\sim\epsilon$.

To leading order in the Navier-Stokes equations, the velocity of the fluid takes the form
of a Poiseuille flow
\begin{align}
&v_{x}=\frac{z^2}{2\mu}\partial_x p+a_xz+b_x,\nonumber \\
&v_{y}=\frac{z^2}{2\mu}\partial_y p+a_yz+b_y,\label{e:vfluid}
\end{align}
where $a_x,b_x,a_y,b_y$ and $p$  do not depend on $z$, but depend on $x$ and $y$.


Following the same lines as in Ref. \cite{LeGoff2014}, we obtain an equation for the evolution of the
 membrane profile 
\begin{align}
\partial_t h=-\nabla_{xy}\left\{
{\cal M}_z(h)\nabla_{xy} f_z 
+ {\cal M}_{\mathbf j}(h) \mathbf{j}
+ {\cal M}_{xy}(h) \mathbf f_{xy}
\right\}
+\frac{\nu}{2}f_z,
\label{eq:evol_h_j_fz_fxy}
\end{align}
where $f_z$ and $\mathbf f_{xy}$ respectively denote the forces acting on the membrane
along $z$ and in the $xy$ plane. In addition, we have defined the total flow
\begin{align}
\mathbf{j}=\int_{-h_0}^{h_0}\!\!\!\! dz\; \mathbf v_{xy}(z).
\label{e:def_j}
\end{align}
Moreover, we have defined the functions
\begin{align}
{\cal M}_z(h)&=
\frac{h_0^3}{24 \mu} \left[ 1- \frac{h^2}{h_0^2}  \right]^3, \nonumber \\
{\cal M}_{\mathbf j}(h)&=
\frac{1}{4} \frac{h}{h_0}\left[ 3 -  \frac{h^2}{h_0^2} \right], \nonumber \\
{\cal M}_{xy}(h)&=
-\frac{h_0^2}{8\mu} \frac{h}{h_0}\left[ 1- \frac{h^2}{h_0^2} \right]^2.
\label{eq:evol_h_j_fz_fxy_mobilities}
\end{align}

Then, using the fluid incompressibility and the boundary condition Eq. \eqref{e:BC_nu} we obtain
\begin{align}
0=\nabla_{xy}\cdot\mathbf{j}+2\nu(\bar p-p_{ext}) 
\label{eq:j_p_equation}
\end{align}
where the average pressure $\bar p=(p_++p_-)/2$  obeys
\begin{align}
2\nabla_{xy}\bar p =-\frac{3\mu}{h_0^3} \mathbf j
+\frac{1}{2}\frac{h}{h_0}\left[ 3 -  \frac{h^2}{h_0^2} \right]\nabla  f_{z}
+\frac{3}{2h_0}\left[ 1 -  \frac{h^2}{h_0^2} \right]\mathbf f_{xy}.
\label{eq:pressure_gradient}
\end{align}

Taking the gradient of Eq. \eqref{eq:j_p_equation} and using \eqref{eq:pressure_gradient},
we obtain an equation for $\mathbf j$ without reference
to pressure
\begin{align}
\nabla_{xy} (\nabla_{xy}\cdot\mathbf{j}) - \frac{3\mu \nu}{h_0^3} \mathbf{j}
=-\frac{\nu}{2}\frac{h}{h_0}\left[ 3- \frac{h^2}{h_0^2} \right] \nabla_{xy} f_z
-\frac{3\nu}{2h_0}\left[1-\frac{h^2}{h_0^2}\right]\mathbf f_{xy}.
\label{eq:j_equation}
\end{align}

We now need to evaluate the forces $f_z$ and $\mathbf f_{xy}$.
In order to do so, we write the variation 
of the energy ${\cal X}$ 
\begin{align}
&\delta{\cal E} + \delta \left( \int \int d {\cal A} \sigma (s_1,s_2) \right) = \int \int d {\cal A} \left[ \kappa \left( \Delta_b {\cal C} +\frac{{\cal C}^3}{2} -2{\cal C}c_G \right) \mathbf{n} \right. \nonumber \\
&-(g^{ij} \partial_{s_i} \sigma \partial_{s_j} \mathbf{r} + \sigma {\cal C} \mathbf{n}) + \nabla {\cal U} (\mathbf{r}) \left. -g^{ij} \partial_{s_i} {\cal U} (\mathbf{r}) \partial_{s_j} \mathbf{r} - {\cal U} (\mathbf{r}) {\cal C} \mathbf{n} \right]\cdot \delta \mathbf{r},
\end{align}
where  $(i,j)$ can take values $(1,2)$, $(s_1,s_2)$ are internal coordinates of membrane, 
$\Delta_b=g^{-1/2}\partial_{s_j}(g^{1/2}g^{ij}\partial_{s_i})$ is the Beltrami laplacian, 
$g^{ij}$ is the inverse metric tensor, 
${\cal C}$ is the mean curvature,
$c_G$ is the Gaussian curvature 
and $\mathbf{n}$ is the unit vector normal to the membrane. 
At each point $\mathbf{r}(s_1,s_2)$ on the membrane surface we can define two tangent vectors
\be
\mathbf{t}_i = \partial_{s_i} \mathbf{r}(s_1,s_2),
\ee
where $i=(1,2)$.

The resulting tangential and normal forces per unit surface are given by \cite{Helfrich1989}
\begin{align}
&f_{t_j}=g^{ij}\partial_{s_i}(\sigma+{\cal U}(\mathbf{r}))-\nabla{\cal U}(\mathbf{r})\cdot \mathbf{t}_j, \nonumber \\
&f_n=-\kappa \left( \Delta_b {\cal C}+\frac{{\cal C}^3}{2}-2{\cal C}c_G \right) + (\sigma + {\cal U}(\mathbf{r})){\cal C}\nonumber\\
&\hspace{1cm}-\nabla {\cal U}(\mathbf{r})\cdot \mathbf{n}.
\end{align}
and in Eq. (\ref{e:membrane_force_equilibrium}) of the main text, 
we use $\mathbf f= f_n \mathbf{n}+ f_{t_i}\mathbf{t}_i$.

We are interested in physical conditions where  
the adhesion potential, the bending rigidity, and tension effects contribute simultaneously
to the normal forces. Hence, we need to require
that ${\cal U}_0/h_0\sim\kappa \epsilon^4\sim \sigma\epsilon^2$. In addition, 
since  the normal force should balance the jump
of pressure $(p_+-p_-)\sim\epsilon^{-1}$
at the membrane from \eqref{e:membrane_force_equilibrium},
we must require $f_z\sim f_n\sim \epsilon^{-1}$.
Combining these two conditions, we obtain that
${\cal U}_0\sim\epsilon^{-1}$, $\kappa\sim\epsilon^{-5}$,
and $\sigma\sim\epsilon^{-3}$ .

Using this scalings, and expanding
$\sigma=\sigma_0+\sigma_1+\sigma_2+...$, with $\sigma_i\sim\epsilon^{-3+i}$,
we find:
\begin{align}
&f_z = - \kappa \Delta_{xy} ^2 h + \sigma_0 \Delta_{xy} h - {\cal U} ^\prime (h)+{\cal O} (1),
\label{eq:fz_lub}
\\
&\mathbf f_{xy}=\nabla_{xy}\sigma_0+\nabla_{xy}\sigma_1+\nabla_{xy}\sigma_2+{\cal O} (\epsilon).
\label{eq:fxy_lub}
\end{align}
Here, we have kept the the sub-dominants contribution in
the expression of the forces in the $xy$ plane for reasons that 
will become clear below.

We now use the membrane area conservation relation Eq. \eqref{e:rho_2D}
which reads to leading order
\begin{align}
\nabla_{xy}\cdot\mathbf v_{xy}(h)=0
\label{e:rho_2Dlead}
\end{align}
where $\mathbf v_{xy}(h)$ is the 2D membrane velocity 
\begin{align}
\mathbf{v}_{xy}=
{\cal N}_{\mathbf j}(h)\mathbf j
+{\cal N}_z(h)\nabla_{xy} f_z
+{\cal N}_{xy}(h)\mathbf f_{xy},
\end{align}
where 
\begin{align}
  {\cal N}_{\mathbf j}({h})&=\frac{3}{4h_0}\left[1-\frac{h^2}{h_0^2}\right],
  \nonumber \\
  {\cal N}_z({h})&=-\frac{h_0^2}{8\mu}\frac{h}{h_0}\left[1-\frac{h^2}{h_0^2}\right]^2,
  \nonumber \\
  {\cal N}_{xy}(h)&=\frac{h_0}{8\mu}\left[1+3\frac{h^2}{h_0^2}\right]\left[1-\frac{h^2}{h_0^2}\right].
\end{align}

Inserting the expression of the forces Eqs. \eqref{eq:fz_lub},\eqref{eq:fxy_lub} in Eq. \eqref{e:rho_2Dlead},
we see that the dominant contribution comes from the term $\sigma_0$
and reads 
\begin{eqnarray}
\nabla_{xy}\cdot\left[{\cal N}_{xy}(h)\nabla_{xy}\sigma_0\right]=0.
\end{eqnarray}
Since periodic boundary conditions are used in this study, 
we conclude that $\sigma_0$ is necessarily a constant in space
from the strong maximum principle \cite{Protter1967}.
Note however, that this is not constant in time. 
To sub-dominant order, the same equations are obtained
and as a consequence $\sigma_1$ is also a constant in space.

To the next order, the membrane area conservation Eq. \eqref{e:rho_2D}
still takes the form Eq. \eqref{e:rho_2Dlead}. Inserting the expression
of the forces Eqs. \eqref{eq:fz_lub},\eqref{eq:fxy_lub} into  Eq. \eqref{e:rho_2Dlead}
then leads to
\begin{align}
0=\nabla_{xy}\cdot\left[
{\cal N}_{\mathbf j}(h)\mathbf j
+{\cal N}_z(h)\nabla_{xy} f_z
+{\cal N}_{xy}(h)\nabla_{xy} \sigma_2
\right].
\label{eq:membrane_local_area_cons_sigma1}
\end{align}
Note also that 
Eq. \eqref{eq:evol_h_j_fz_fxy} can be written as
\begin{align}
\partial_t h=-\nabla_{xy}\left\{
\frac{h_0^3}{24 \mu} \left[ 1- \frac{h^2}{h_0^2}  \right]^3\nabla_{xy}f_z 
+\frac{1}{4} \frac{h}{h_0}\left[ 3 -  \frac{h^2}{h_0^2} \right] \mathbf{j} \right. \nonumber \\
- \left. \frac{h_0^2}{8\mu} \frac{h}{h_0}\left[ 1- \frac{h^2}{h_0^2} \right]^2 \mathbf \nabla_{xy}\sigma_2
\right\}
+\frac{\nu}{2}f_z.
\label{eq:evol_h_j_fz_sigma1}
\end{align}
Solving this latter equation requires the knowledge of $\sigma_0(t)$, $\mathbf j(x,y,t)$, and
$\sigma_2(x,y,t)$ at each time.

The expression of $\sigma_0$ is found from 
the global conservation of the excess area Eq. \eqref{e:excess_area_normalized}.
Indeed, from $\partial _t \Delta {\cal A}=0$,
using periodic boundary conditions, we have
\begin{align}
0= \int \int dx dy \; \partial_t h \;\Delta h
\label{eq:area_conservation_physvar}
\end{align}
which provides an expression for $\sigma_0$
as a function of $\mathbf j$ and $\mathbf \sigma_2$.

Then, the two equations \eqref{eq:membrane_local_area_cons_sigma1}
 and \eqref{eq:j_equation} provide a 
 linear system of differential equations with space-dependent coefficients 
 and no time-derivative,
 which must be solved at each time to obtain $\mathbf j$ and $\mathbf \sigma_2$.

\subsection{Limit of very permeable walls}

In the limit of very permeable walls,
the only term that survives in Eq. \eqref{eq:evol_h_j_fz_sigma1}
is the last one $(\nu/2)f_z$. Hence, the 
sub-dominant and space-dependent tension $\sigma_2(x,y,t)$ is irrelevant.
The leading-order space-independent tension $\sigma_0(t)$
calculated from Eq. \eqref{eq:area_conservation_physvar}
leads to Eq. \eqref{eq:sigmapermeable}.

\subsection{Limit of impermeable walls}

In this case, the last term in Eq. \eqref{eq:evol_h_j_fz_sigma1} 
proportional to $\nu$ is negligible.

In addition, the total flow $\mathbf j$ obeys a simplified
equation as compared to \eqref{eq:j_equation}:
\begin{align}
\nabla_{xy}\cdot\mathbf j=0.
\end{align}
This equation is  
expressing that the total mass of liquid is then locally conserved.
Taking the curl of Eq. \eqref{eq:pressure_gradient}, we obtain a second
equation 
\begin{align}
&\frac{3\mu}{h_0^3}\nabla_{xy}\times\mathbf j
\nonumber \\
&=\frac{3}{2}\left[1-\frac{h^2}{h_0^2}\right]\frac{\nabla h \times\nabla f_{z}}{h_0}
-3\frac{h}{h_0}\frac{\nabla h \times\nabla \sigma_2}{h_0^2}\,.
\end{align}
These two scalar equations, together with Eq. \eqref{eq:membrane_local_area_cons_sigma1}
and suitable boundary conditions allow one to determine $\mathbf j$ and $\sigma_2$.

 


In the main text, we present simulations for a simplified conserved
model, where $\sigma_2$ and $\mathbf j$ are neglected. 
Then, using  Eq. \eqref{eq:area_conservation_physvar}
leads to the expression of $\sigma_0$ given in Eq. \eqref{eq:sigmaimpermeable}.



\section{Decrease of the total energy}
\label{b:energyrelaxation}

In this section we consider the general case of a membrane of height $h$ with the energy
\begin{align}
{\cal E} = \int d {\cal A} g,
\end{align}
where $g$ is an energy density depending on $h$ and its derivatives,
and the related force
\begin{align}
f=\frac{\delta {\cal E}}{\delta h} .
\end{align}
The dynamics is ruled by one of the following equations
\begin{align}
\partial_th&=-f
+\sigma_0\Delta h,
\label{eq:gen_very_perm}\\
\partial_th&=\nabla\cdot \left[
{\cal M}(h)
\nabla\left(f
-\sigma_0\Delta h\right)
\right]
\label{eq:gen_imperm},
\end{align}
where ${\cal M}(h)$ is a 
positive nonlinear mobility depending on $h$,
and the space-independent tension $\sigma_0$ enforces the total area conservation of the membrane (see Appendix \ref{a:lubrication}). In the current study, these two situations correspond respectively to the limits of very permeable walls and impermeable walls of Eq. \eqref{eq:evol_h_j_fz_sigma1}. 

As discussed in Appendix \ref{a:lubrication},
area conservation is enforced by the relation
\begin{align}
\sigma_0=\frac{\int d{\cal A}\; f\Delta h }{\int d{\cal A}\; (\Delta h)^2 }.
\end{align}

In the limit of very permeable walls described by Eq. \eqref{eq:gen_very_perm}, we have
\begin{align}
\partial_t{\cal E} &= \int d {\cal A} f\partial_th,\nonumber\\
&= -\int d {\cal A} f^2
+\int d {\cal A} f\sigma_0\Delta h
\nonumber\\
&=
-\int d {\cal A} f^2
+\frac{\left(\int d{\cal A}\; f\Delta h \right)^2}{\int d{\cal A}\; (\Delta h)^2 }
\end{align}
Then, from the Schwarz inequality we have
\begin{align}
 \left(\int d {\cal A} f \Delta h\right)^2
 \leq 
 \int d {\cal A} (\Delta h)^2 \int d {\cal A} f^2
 \end{align}
 leading to
$\partial_t{\cal E}\leq
0.$

In the opposite limit of impermeable walls described by Eq. \eqref{eq:gen_imperm}, we now have
area conservation is imposed via the relation
\begin{align}
\sigma_0=\frac{\int d {\cal A} \nabla\cdot \left[
{\cal M}(h)
\nabla f
\right]\Delta h}{\int d {\cal A} \nabla\cdot \left[
{\cal M}(h)
\nabla\left(\Delta h\right)
\right]\Delta h}.
\end{align}
This expression is identical to that reported in the main text in Eq. \eqref{eq:sigmaimpermeable}.
Using integration by parts and periodic boundary conditions, this leads to
\begin{align}
\partial_t{\cal E}=&-\int d {\cal A} 
{\cal M}(h)
(\nabla f)^2\nonumber\\
&+\frac{\left[\int d {\cal A} 
{\cal M}(h)
\nabla f
\cdot\nabla\left(\Delta h\right)\right]^2}{\int d {\cal A} {\cal M}(h)\left[\nabla\left(\Delta h\right)\right]^2}.
\end{align}
Using once again the Schwarz inequality, we find
$\partial_t{\cal E}\leq
0.$






\section{Numerical methods}

\subsection{Area conservation}
\label{a:area_conservation}

We choose a numerical method to determine $\Sigma$
which minimises the error on area conservation.
In practice, we impose that $\Delta A_{T+dT}=\Delta A_T$.
Since both in the conserved and non-conserved
regimes the quantity $H_{T+dT}$ is linear in $\Sigma$, 
and since $\Delta A_{T+dT}$ is quadratic in $H_{T+dT}$, the
conservation of the excess area implies
the solution of a quadratic equation for $\Sigma$.
This quadratic equation has two solutions.
We choose the physically relevant solution,
which is the one which is the closest to
the value of $\Sigma$ calculated via a direct
estimate Eq. (\ref{eq:sigmapermeable_norm}) using $H=H_T$.

Our scheme can be seen
as a specific discretization of 
Eq. (\ref{eq:sigmapermeable_norm}) using 
a combination of $H_T$ and $H_{T+dT}$.
The resulting variations in $\Delta A$  are $\sim 10^{-10}$.

\subsection{Initial conditions}
\label{a:initial_conditions}

Our simulation scheme with area conservation
requires a smooth initial condition for the 
excess area to be well defined.
We have generated random smooth initial conditions with different 
excess area $\Delta A^*$ using two different methods. 
The first method uses the solution of the  the  
time-dependent Ginzburg-Landau (TDGL) equation 
\be
\partial _T H = w^2 \Delta H - U ^\prime (H),
\ee
using an explicit scheme with finite-differences and random initial conditions.
For each value of $w$, we take the membrane profile corresponding
to the maximum value of $\Delta A^*$ 
as an initial condition for our simulation.
To verify that this procedure does not affect the 
dynamics, we have repeated it  solving the TDGL4 equation
\be
\partial _T H = - w^4 \Delta ^2 H - U ^\prime (H).
\ee 
The final results were similar when considering 
similar initial $\Delta A^*$.

\subsection{Short-range repulsion near the walls}
\label{a:short_range_repulsion}

Since the membrane
does not approach the walls too much
for $\Delta A^* < \Delta A^*_{nl}$,
we have used the interaction potential 
without $U_d$.

For $\Delta A^* \geq \Delta A^*_{nl} = (5.53 \pm 0.15) \cdot 10^{-2}$, 
we use the double well potential with a short-range repulsion
potential $U_d$ near the wall to prevent the membrane height 
from crossing the walls at $Z=\pm 1$, with $U_0=1$ and $d=0.01$. 
This requires a smaller $dT$ for numerical stability.

\subsection{Evaluation of the lengthscales in simulations}
\label{a:lengthscales}

We calculate the typical lengths of flat
domains $\lambda_{flat}$ and wrinkle domains $\lambda_{wr}$  
from the image of the membrane profile. 
First, we remove the localised structures. 
Next, we label positive regions ($H>0$) 
by 2 and negative regions ($H<0$) by 1. 
Then, we erode the boundaries of domains 1 and 2 with 
a disk of radius $\theta=4$. This procedure
removes domain walls. In addition, since this radius is larger
than the half-width of the wrinkles, the
procedure also subtracts the wrinkles regions
from the zones with labels 1 or 2.
The eroded zones are labeled by 0. 
The typical length of flat domains is given by
\be
\label{eq:lambdaflat}
{\lambda}_{flat}=\frac{A_{flat}}{L_{flat}},
\ee
where $A_{flat}$ is the total area formed by domains 1 and 2. 
$L_{flat}$ is the total length of the boundaries of domains 1 and 2 and is calculated using the Cauchy-Crofton formula \cite{CauchyCrofton2007} with two perpendicular and two diagonal sets of parallel lines forming a grid.

Next, we erode the boundaries of domains 0 with a disk of radius $R=8$.
This procedure removes the domain walls between the flat domains
 1 and 2 from domains 0. 
 We calculate the typical length of wrinkles domains $\lambda_{wr}$ 
 from the remaining  0-regions by
\be
\lambda_{wr}=\frac{A_{wr}}{L_{wr}},
\label{eq:lambda_wr_num}
\ee
where $A_{wr}$ is the total area formed by 0-domains. 
$L_{wr}$ is the total length of the boundaries of the domains 0 
and is again calculated using the Cauchy-Crofton formula \cite{CauchyCrofton2007}.
The expressions (\ref{eq:lambdaflat},\ref{eq:lambda_wr_num}) 
are used in the main text in regime B.

In regime A, we evaluate $\bar{\lambda}_{flat}$
defined from Eq.~\eqref{e:bar_lambda_flat}.
In order to determine this quantity, we first notice that after eroding flat domains with the disc, 
the total length of the boundaries of flat domains $L_{DW}$ is doubled, thus
$L_{flat} \approx 2L_{DW}$.
Moreover,   
during the disc erosion step 
the typical domain area is reduced by an amount $2L_{DW}\theta$, 
where $\theta=4$ is the erosion disc radius.
This leads to $A_{syst}=A_{flat}+2\theta L_{DW}$. 
Combining these relations, we obtain
\begin{align}
\bar\lambda_{flat}=\frac{A_{syst}}{L_{DW}}\approx 2(\lambda_{flat}+\theta)
\end{align}
which is used in the main text.

\section{Motion by curvature in the large permeability limit}
\label{a:motion _by_curvature}

Consider a domain wall between two opposite flat domains.
To leading order for small domain wall curvature $K$, the Laplacian
operator can be expanded as
$\Delta \approx \partial _{\zeta \zeta} + K \partial _\zeta$,
where $\zeta$ is a local coordinate along the normal to the domain wall. 
Expanding Eq. (\ref{eq:hpermeable_norm}), we obtain
\be
\label{eq:vn1}
-V_n \partial_\zeta H \approx -\partial^4_\zeta H - 2 K \partial^3_\zeta H + \Sigma \partial^2_{\zeta} H + \Sigma K \partial_\zeta H - U^\prime (H),
\ee
where $V_n$ is the normal front velocity. 
Multiplying both sides of Eq. \eqref{eq:vn1} by 
$\partial_\zeta H$ and integrating with respect to $\zeta$ we find
\begin{align}
V_n=-\frac{1}{\alpha_{DW}} \Big( [U_0]^+_- + K \xi_{DW} \Big).
\end{align}
where
\begin{align}
U_0=U(H)+\frac{1}{2}\partial_{\zeta\zeta}(\partial_\zeta H)^2-\frac{3}{2}(\partial_{\zeta\zeta}H)^2
-\frac{\Sigma}{2}(\partial_{\zeta}H)^2
\label{e:U_0_interactions}
\end{align}
accounts for interactions between domain walls, 
as discussed in Ref. \cite{LeGoff2015StatMech}.
Here, as in the main text $\alpha_{DW}=(1/2)\int d\zeta (\partial_\zeta H)^2$, 
and $[\;]^+_-$ denotes the difference between 
the value of a given quantity in the adjacent adhesion domains
on both sides of the domain wall. Moreover, 
$\xi_{DW}$ is the $\Xi$-energy of a flat domain wall per unit length 
\be
\xi_{DW} = \int d\zeta \Big\{ 
\frac{1}{2}(\partial ^2_\zeta H)^2 + U(H) 
+ \frac{\Sigma}{2}  (\partial_\zeta H)^2 \Big\}.
\ee
To leading order, area conservation imposes
\begin{align}
0&=\partial_T \Big( L_{DW} \int d\zeta (\partial_\zeta H)^2 \Big)
\nonumber \\
&=-\int d\zeta \Big( [U_0]^+_- K 
+ \xi_{DW} K^2 \Big) 
+ L_{DW} \partial_T\alpha_{DW}.
\end{align}
Using the chain rule 
$\partial_T \alpha_{DW}=\partial_T \Sigma\, \partial_\Sigma \alpha_{DW}$,
we find an evolution equation for the tension Eq. (\ref{e:dSdT}).




\section{Sine-profile ansatz in Regime C}
\label{a:sine_ansatz_regimeC}

Here, we use a sine-profile ansatz
\begin{align}
H(\zeta)=a \cos( q \zeta)
\end{align}
where $q=2\pi/\lambda_{1roll}$
to model the wrinkle phase.
In normalised coordinates, the $\Xi$-energy per unit length of wrinkle then reads
\begin{align}
\xi_{1roll}=\frac{\pi}{q}
\left[
\left(\frac{q^4}{2}+\frac{\Sigma q^2}{2}-\frac{ H_m^2}{2}\right)a^2
+\frac{3a^4}{16} 
+\frac{H_m^4}{4}
\right].
\end{align}

Minimizing the energy density $\xi_{1roll}/\lambda_{1roll}$ with respect to $a$ and $q$,
we find
\begin{align}
\Delta A^* &=\frac{-\Sigma H_m^2}{6} 
\left[1+\left(\frac{-\Sigma}{2H_m}\right)^2\right],
\\
\langle H^2\rangle &=\frac{2H_m^2}{3}
\left[1+\left(\frac{-\Sigma}{2H_m}\right)^2\right],
\\
\lambda_{1roll}^2 &=-8\frac{\pi^2}{ \Sigma}.
\end{align}
The value of $\Sigma$ is obtained from the solution of the first
equation. Then, $\langle H^2\rangle$ and $\Sigma$ are obtained
from the two other equations.
The above results correspond to the non-contact regime
where $\Delta A^*<\Delta A^*_{wc}$, and $a<1$.
The value of the critical tension $\Delta A^*_{wc}$ is obtained
from the condition $a=2^{1/2} \langle H^2\rangle^{1/2}=1$.
The expression of $\Delta A^*_{wc}$ is provided in Eq. \eqref{eq:Delta_A*_st}.

In wall-contact regime for $\Delta A^*>\Delta A^*_{wc}$,
we set $a=1$, and minimise the energy density $\xi_{1roll}/\lambda_{1roll}$ 
with respect to $q$ only. This leads to
\begin{align}
\langle H^2\rangle &=\frac{1}{2},
\\
\Sigma &=-8\Delta A^* ,
\\
\lambda_{1roll}&=\frac{\pi}{\Delta A^{*\,1/2}}.
\end{align}




\end{appendix}

\bibliography{adhesionref}
\bibliographystyle{rsc} 

\end{document}